\documentclass{revtex4}

\usepackage{amssymb}
\usepackage{amsmath}

\usepackage{color}
\definecolor{red}{rgb}{1,0,0}

\definecolor{green}{rgb}{0,1,1}

\def\giorno{October 15, 2015}

\def\ri{\mathrm{i}\,}
\def\rj{\mathrm{j}\,}
\def\rk{\mathrm{k}\,}
\def\R{\mathbf{R}}
\def\be{\begin{equation}}
\def\ee{\end{equation}}

\def\Jb{{\bf J}}
\def\w{\wedge}
\def\d{{\mathrm d}}

\def\V{V}
\def\L{M}

\def\what{\widehat}
\def\wt{\widetilde}
\def\EOR{ \hfill $\odot$}
\def\EOP{ \hfill $\triangle$}

\def\eqref#1{(\ref{#1})}

\def\quaternionic{quaternionic }
\def\HK{hyperk\"ahler }
\def\hK{hyperk\"ahler }

\def\HS{hypersymplectic }

\begin{document}

\title{Structure preserving transformations in \HK Euclidean spaces}

\author{G. Gaeta$^1$\footnote{E-mail: giuseppe.gaeta@unimi.it}
and M. A. Rodr\'{\i}guez$^2$\footnote{E-mail: rodrigue@fis.ucm.es}}

\affiliation{$^1$Dipartimento di Matematica,
Universit\`a degli Studi di Milano,
via Saldini 50, 20133 Milano (Italy)\\
$^2$Departamento de F\'{\i}sica Te\'orica II,
Universidad Complutense, 28040 Madrid (Spain)}

\date{\giorno}

\begin{abstract}
The definition and structure of \HK structure preserving
transformations (invariance group) for \quaternionic structures
have been recently studied and some preliminary results on the
Euclidean case discussed. In this work we present the whole
structure of the invariance Lie algebra in the Euclidean case  for
any dimension.
\end{abstract}

\maketitle
\section{Introduction}

Hyperk\"ahler manifolds are a traditional field of study for
geometers  \cite{AM96,At90,Ca79,Ca80} but it has became also of
interest to Theoretical Physics, in particular after the
pioneering work of Atiyah, Hitchin and collaborators
\cite{At79,AH88,HK87}; see also the bibliography in \cite{Du10} and
\cite{GM94,MP01} for more recent contributions.

Motivated mainly by our study of hyperhamiltonian dynamics
\cite{GM02,MT03} and its applications in Physics \cite{GR08,GR12}
(and integrable systems \cite{Ga11,GM03a}), we are interested
in  canonical transformations in hyperhamiltonian dynamics; this also
led to investigating transformations in \HK manifolds which preserve
the \HK structure;  this is a topic which of course has been already considered
by Differential Geometry -- albeit from a geometric rather than dynamical point
of view \cite{Jo00}. The requirement of strictly preserving
each of the complex structures (which leads to tri-holomorphic
maps) is exceedingly restrictive, and one should instead focus on
more relaxed requirements.

In standard Hamilton dynamics or symplectic geometry one requires
the preservation of the symplectic structure (i.e. of the
symplectic form $\omega$). In the framework of \HK manifolds we
are in the presence of three symplectic structures $\omega_\alpha$
($\alpha=1,2,3$), associated via the K\"ahler relation $\omega_\alpha = (J_\alpha . \, , \, . )$ to the
complex structures $J_\alpha$ (and to the Riemannian metric $g$)
defined on the manifold $M$. Under many aspects (and in particular
for hyperhamiltonian dynamics) it is natural to consider a set $\{
\wt{\omega}_\alpha \}$ obtained as $\wt{\omega}_\alpha = R_{\alpha
\beta} \omega_\beta$, with $R$ a matrix in SO(3), as {\it
equivalent} to the set $\{ \omega_\alpha \}$. Thus one considers
transformations in $M$ which preserve the metric and which map the
set $\omega_\alpha$ to an equivalent one; these are called {\it
hypersymplectic} or {\it quaternionic\/}. We stress these are of
interest not only for hyperhamiltonian dynamics but for \HK geometry
as well: the invariance group of quaternionic structures has been
studied and identified in the Differential Geometry literature devoted
to \hK and quaternionic manifolds \cite{Jo00}, based on a rather abstract
approach; on the other hand, our discussion will be based on very explicit
linear algebra construction and some standard (classification) theory of Lie
algebras. We trust this approach may be more familiar to physicists, and we
believe it is worth having a completely explicit discussion.

In this paper, we investigate the (connected component of the)
group of \quaternionic transformations for Euclidean spaces $\R^{4
n}$ of arbitrary dimension $4 n$; this is of course much simpler
than the general case but is a necessary first step before dealing
with more complex situations.
It turns out that in this case one is able to provide a fairly
complete characterization of the Lie algebra of this group, also
called the {\it invariance algebra} $\mathfrak{L}_n$ below, in
arbitrary dimension.

We show (Theorem 3) by a completely explicit
procedure (based on standard linear algebra and in which the main difficulty is that
of having a convenient notation, plus some general results from the theory of Lie algebras), that $\mathfrak{L}_n = \mathfrak{su}(2) \oplus \mathfrak{sp}(n)$.
We also show (Theorem 5),  again in a fully transparent way, that the ``strong invariance algebra'', leaving each of the $\omega_\alpha$ invariant, is $\mathfrak{g}_n = \mathfrak{sp}(n)$.

 As already mentioned, these results are not new {\it in se}, being known since some time in the differential geometric literature \cite{Jo00}; but they were obtained in a rather abstract way, while the derivation we provide here is fully explicit and based on standard linear algebra.

These results are of course also in agreement with Berger's list of holonomy
groups for Riemannian manifolds \cite{Be55} and further research on this topic
(see \cite{Jo00} for a comprehensive exposition of this subject; see also \cite{GM94,MP01}).
In fact, the structure group of the manifolds under study contains the holonomy groups, and -- as it should be expected since there are no other additional structure involved -- it turns out they coincide.

The work presented in this paper contains a detailed description of the representations of these groups appearing in the structure of the holonomy groups, using the standard representations of the quaternionic structure in $\R^4$.

Note that some ambiguity is present in the literature concerning
the notation for symplectic groups; for us $\mathrm{Sp}(n)$ will
be the set of $(2n \times 2n) $ (complex) unitary symplectic
matrices (thus with real representation of dimension $4n$), with
Lie algebra $\mathfrak{sp} (n) \subset \mathrm{Mat} (2n,{\bf C})
\simeq \mathrm{Mat} (4n,\R)$.

 Finally, we would like to briefly stress the {\it physical} relevance of
the flat case. This is due not only to the fact the Dirac equation is set in
flat Minkowski space (which would maybe suffice by itself), but also to the fact
that most of the physically relevant nontrivial \hK manifolds are obtained from
higher dimensional Euclidean $\R^{4n}$ manifolds (with standard \hK structure) via
the moment map construction pioneered by Hitchin {\it et al.} \cite{HK87}; thus e.g.
the \hK structure (beside of course the metric) for the Taub-NUT manifold \cite{NT63,Ta51}
can be built explicitly starting from those in $\R^8$ \cite{GR12}. Thus, albeit maybe not
so interesting for Geometry, the Euclidean case has a substantial relevance for Physics,
and we believe it is worth having a fully explicit discussion of the invariance group for
\hK structures in Euclidean $\R^{4n}$ spaces.

The {\it plan of the paper} is as follows. In Section
\ref{sec:background} we will recall some basic facts about \HK
manifolds and maps defined on it, also to set our general
notation; here we will also discuss some facts about orientation
in \HK manifolds, to be used in the following. In Section
\ref{sec:euclidean} we will specialize to the simplest \HK
manifolds, i.e. spaces $\R^{4n}$ with Euclidean metric and three
complex structures satisfying the quaternionic relations (these
define symplectic structures via the K\"ahler relation); we argue
that in this framework we can reduce to a simple setting, i.e. to
{\it standard \HK structures}, and we can thus reduce to a problem
defined in terms of these standard structures (which may still
have different orientations). In this section we will also set the
stage to discuss hypersymplectic maps at the infinitesimal level.
After this general discussion, we will tackle the simplest case
$n=1$ in Section \ref{sec:R4}, showing how the basic problem can
be solved in very explicit terms. Higher dimensional cases are
considerably more involved; in order to help the reader
familiarize with the tools needed to deal with the general case we
give, in Section \ref{sec:R8}, a separate discussion of the $n=2$
case: this presents the difficulties of the general case, but it
is still possible to follow the problem, and its combinatorial
aspects, in a nearly explicit manner. Here four different
orientations are possible for the \HK structure (including
standard ones), and the discussion of Section \ref{sec:background}
comes to our help in order not to have to discuss each of them
separately. In Section \ref{sec:R4n} we discuss the
general case, using the tools developed for $n=1$ and $n=2$ as
well as some general results from Lie algebra theory, and, finally, we draw our conclusions in
Section \ref{sec:conclusions}. Some
technical details of the $n=2$ and general $n$ case discussion and
computations are confined to the two appendices. The symbols
$\triangle$ and $\odot$ signal, respectively, the end of proofs
and remarks.

\section{Basic notions}
\label{sec:background}

We will start by recalling the basic notions needed for our
discussion; that is, we will recall what are \HK manifolds and
which are the characteristics required to a map on a \HK manifold
to consider it as preserving the structure on it.

In the following $I_k$ will denote the $k$-dimensional identity
matrix; we will also denote by $\mathcal{M}_k$ the set of
$k$-dimensional real matrices, i.e. $\mathcal{M}_k := \mathrm{Mat}
(k,\R)$.

\subsection{Hyperk\"ahler manifolds}

We will give only the basic definition of \HK manifold. For
details on \HK manifolds, see e.g.
\cite{AM96,At79,At90,AH88,Du10}.

\medskip\noindent {\bf Definition 1.} {\it A {\it \HK manifold }
$(\V,g;\Jb)$ is a real smooth orientable Riemannian manifold
$(\V,g)$ of dimension $m = 4n$ equipped with an ordered triple
$\Jb = \{ J_1,J_2,J_3 \}$ of orthogonal almost-complex structures which are
covariantly constant under the Levi-Civita connection, $\nabla
J_\alpha = 0$; and satisfy the quaternionic relations \be
\label{quatcond} J_\alpha \, J_\beta \ = \ \epsilon_{\alpha \beta
\gamma} \, J_\gamma \ - \ \delta_{\alpha \beta} \, I \ . \ee }
\bigskip

The requirement $\nabla J_\alpha = 0$ implies that the $J_\alpha$ are
actually complex structures on $(\V,g)$, due to the
Newlander-Nirenberg theorem \cite{NN}.

The ordered triple $\Jb = \{J_1,J_2,J_3\}$ will also be called a
{\it \HK structure} on $\V$; thus a \HK
manifold is an orientable smooth manifold $\V$ equipped with a
Riemannian metric $g$ and with a \HK structure invariant
under the associated Levi-Civita connection $\nabla$.

\medskip\noindent {\bf Definition 2.} {\it Let $\Jb$ and
$\what{\Jb}$ be different \HK structures on the same Riemannian
manifold $(\V,g)$; if each of them can be expressed in terms of
the other, \be\label{equivHK} \what{J}_\alpha \ = \ \sum_{\beta =
1}^3 \, r_{\alpha \beta} \, J_\beta\ , \ee the two structures are
said to be {\rm equivalent}. An equivalence class of \HK
structures on $(\V,g)$ is said to be a {\rm quaternionic
structure} on $(\V,g)$.}
\bigskip

It should be stressed that since both the $J$ and the $\what{J}$
satisfy the quaternionic relations \eqref{quatcond}, necessarily
the matrix $R$ with entries $r_{\alpha \beta}$ in \eqref{equivHK}
belongs to the Lie group $\mathrm{SO}(3)$. Moreover, as both $J$ and
$\what{J}$ are covariantly constant, it follows that $\nabla R =
0$ as well.

\medskip\noindent {\bf Remark 1.} Hyperk\"ahler structures related by
linear transformations such as those considered in Definition 2
should in many aspects be seen as substantially equivalent (hence
the notion of equivalent structures). In this sense, the relevant
structure is the quaternionic one; we will take this into account
when looking for structure-preserving transformations on $\V$.
\hfill\EOR
\bigskip

If we define local coordinates in $\V$, the (1,1) tensors
$J_\alpha$ are represented by matrices (which we denote again by
$J_\alpha$ with a standard abuse of notation), and the
quaternionic relation \eqref{quatcond} holds between such
matrices.

Associated to the metric and each of the complex structures
$J_{\alpha}$ we can construct three symplectic forms
$\omega_{\alpha}$ by the K\"ahler relation; then
$(\V,g,\omega_\alpha)$ is a K\"ahler manifold for any $\alpha$. We
also say that $(\V,g;\omega_1,\omega_2,\omega_3)$ is a {\it
hypersymplectic manifold}.

\medskip\noindent {\bf Remark 2.} Thus a \HS manifold is an
orientable smooth manifold $\V$ of dimension $4 n$, equipped with
a Riemannian metric $g$ and an ordered triple of covariantly
constant symplectic forms $\omega_\alpha$, such that the complex
structures $J_\alpha$ obtained from these via the K\"ahler
relation obey the quaternionic relations \eqref{quatcond}. Note
that here, differently from the standard symplectic case, the
metric plays a key role through the K\"ahler relation. \EOR

\medskip\noindent {\bf Remark 3.} The notion of equivalent \HK
structures induces naturally a notion of equivalent \HS
structures: two \HS structures are equivalent if the complex
structures $\Jb$ and $\what{\Jb}$ they induce via the K\"ahler
relation are equivalent, i.e. satisfy \eqref{equivHK}. \EOR
\bigskip

In local coordinates on $\V$, these symplectic forms are written as
\be \omega_\alpha \ = \ \frac{1}{2} \, (K_\alpha)_{ij} \, \d x^i
\w \d x^j \ , \ee with $K_\alpha = g J_\alpha$; it follows from
Definition 1 that the $K_\alpha$ are covariantly constant under
the Levi-Civita connection and (see \eqref{quatcond} above) that
they satisfy \be \label{quatcondsymp} K_\alpha \, g^{-1} \,
K_\beta \ = \ \epsilon_{\alpha \beta \gamma} \, K_\gamma \ - \
\delta_{\alpha \beta} \, I \ . \ee

\subsection{Maps on \HK manifolds}

We would now like to characterize the maps $\varphi : \V \to \V$
which leave invariant the \HK structure, or at least the
equivalence class of \HK structures discussed above,
i.e. the quaternionic structure on $(\V,g)$.

If $\varphi : \V \to \V$ is an arbitrary smooth map in $\V$, the
\HK structure will change according to the rule of
transformations of $(1,1)$ tensors (which amounts to a
conjugation), i.e.
$$ J_{\alpha} \ \to \ \wt{J}_{\alpha} \ :=
\Lambda \, J_{\alpha} \, \Lambda^{-1} \ , $$
where $\Lambda$ is the Jacobian of $\varphi$.

\medskip\noindent {\bf Definition 4.} {\it The map $\varphi : \V
\to \V$ is {\rm strongly \HK} for $(\V,g,\Jb )$ if it preserves
both the metric $g$ and the \HK structure $\Jb$.}
\bigskip

It is obvious that the set of strongly \HK maps for $(\V,g,\Jb )$
is a group. Such maps are also called {\it tri-holomorphic}, as
they are holomorphic for each of the three complex structures
$J_\alpha$. The group of strongly \HK maps on $(\V,g;\Jb)$ will be
called the strong \HK group on $(\V,g;\Jb)$; or for short the {\it
strong invariance group of $\V$}. Correspondingly, its elements
will be called, with an abuse of language, {\it strong invariance
maps}.

\medskip\noindent {\bf Definition 5} {\it The map $\varphi : \V \to
\V$ is {\rm hyperk\"ahler} for $(\V,g,\Jb )$ if it preserves the
metric and maps the \HK structure into an equivalent one. In this
case it is also said to be {\rm quaternionic}, as it preserves the
quaternionic structure on $(\V,g)$.}
\bigskip

Here again it is obvious that the set of \HK maps for
$(\V,g,\Jb )$ is a group. This will be called the \HK group on
$(\V,g;\Jb)$; or for short the {\it invariance group of $\V$}.
Correspondingly, its elements will be called, with an abuse of
language, {\it invariance maps}. Any strong invariance map is also
an invariance map; the set of maps which are \HK but not strongly
\HK will also be denoted as {\it regular invariance maps}.

\medskip\noindent {\bf Remark 4.} It is clear that, by the
correspondence mentioned at the end of the previous subsection
(and based simply on the K\"ahler relation), the maps preserving
the \HK structure will also preserve the \HS one, and those
mapping the \HK structure into an equivalent one will also maps
the \HS structure into an equivalent one. Thus it would also be
legitimate to denote the maps and groups identified in the
Definitions 4 and respectively 5 above as strongly \HS and
respectively \HS ones; the groups will correspondingly be called
the {\it strong \HS group} and the {\it \HS group}. \EOR

\subsection{Orientation}

As recalled above, a \HK manifold is orientable; we can thus
consider in particular orientation-switching maps $\mathcal{P}$,
obviously satisfying $\mathcal{P}^2 = I$. Under such a map the \HK
structure will not be preserved, but will be mapped to a
(non-equivalent) {\it dual} one \cite{GR14a}; note that the
Riemannian metric can instead be invariant under such an
orientation-switching map (this will in particular be the case for
the Euclidean metric for any orthogonal $\mathcal{P}$). From now
on we will only consider maps $\mathcal{P}$ preserving the metric.

It is quite obvious that \HK structures which are dual to each
other (we refer to these as a {\it dual pair}) are strongly
related; it also turns out that in some physical applications of
hyperhamiltonian dynamics (in particular, in the description of
the Dirac equation in hyperhamiltonian terms \cite{GR08}) one
needs both elements of a dual pair.

When we work in the symplectic framework, so that the \HK
structure corresponds to a triple of symplectic structures $\{
\omega_\alpha \}$, the action of the map $\mathcal{P}$ on these is
simply given by the pull-back. This induces a conjugation between
the $\omega_\alpha$ and the dual ones, $\wt{\omega}_\alpha =
\mathcal{P}^* \omega_\alpha$, and hence between a \HK structure
and its dual one. This shows that \HK structures related by such
an orientation switch are conjugated. (Representing the forms
$\omega_\alpha$ (and the complex structures $J_\alpha$) in
coordinates, the conjugation is described by the action of a
matrix $P$ which is orthogonal with respect to the metric.)

We conclude that \HK structures related by such a map will be
invariant under  isomorphic groups of transformations.

In the following we will have to consider spaces $\R^{4n}$ and the
possibility to independently switch orientation in the $\R^4$
subspaces on which the $\omega_\alpha \w \omega_\alpha$ give a
volume form; the same considerations presented above will also
hold for the restriction of the \HK structure to each of these
subspaces, and this will be rather useful to simplify our
computations.

\section{Euclidean spaces}
\label{sec:euclidean}

In this note we are concerned with the simplest occurrence of \HK
manifolds, i.e. Euclidean spaces. We will thus specialize our
general notions and  discussion to this specific case.

\subsection{Hyperk\"ahler structures}

The simplest example of a \HK manifold is $\R^{4n}$ with the
Euclidean metric, equipped with the {\it standard \HK structures}
detailed below (these will play a role in our general discussion).
It should be noted that, since the metric is here Euclidean, the
covariant derivative is the usual derivative (the Levi-Civita
connection is trivial). Then $\partial_{x_i} J(x) = 0$ for $i=1,...,4n$,
and the \HK structure is actually constant.

The standard positively or negatively oriented standard \HK
structures in $\R^4$ are given respectively by
\be  Y_1
=  \begin{pmatrix}  0&1&0&0\\ -1&0&0&0\\ 0&0&0&1\\
0&0&-1&0 \end{pmatrix} , \quad Y_2 = \begin{pmatrix}
0&0&0&1\\ 0&0&1&0\\ 0&-1&0&0\\ -1&0&0&0 \end{pmatrix},\quad
Y_3 =\begin{pmatrix} 0&0&1&0\\ 0&0&0&-1\\ -1&0&0&0\\
0&1&0&0\end{pmatrix}, \ee \be
 \what{Y}_1 = \begin{pmatrix}0&0&1&0\\ 0&0&0&1\\
-1&0&0&0\\ 0&-1&0&0\end{pmatrix}, \quad
\what{Y}_2 = \begin{pmatrix}0&0&0&-1\\
0&0&1&0\\Ê0&-1&0&0\\ 1&0&0&0\end{pmatrix}, \quad \what{Y}_3 =
\begin{pmatrix}0&-1&0&0\\ 1&0&0&0\\ 0&0&0&1\\
0&0&-1&0\end{pmatrix}. \ee

\medskip\noindent {\bf Remark 5.} The qualification on orientation
follows from considering the associated symplectic structures
$\omega_\alpha$ and $\what{\omega}_\alpha$; with $\Omega$ the
standard volume form on $\R^4$, one has $(1/2) (\omega_\alpha \w
\omega_\alpha) = \Omega$ and $(1/2) (\what{\omega}_\alpha \w
\what{\omega}_\alpha) = - \Omega$, for all $\alpha$. \EOR
\bigskip

In the four-dimensional case ($n=1$), it is easily checked that
any constant \HK structure $J_{\alpha}$, i.e. any set of constant
skew-symmetric matrices satisfying the quaternionic relations
\eqref{quatcond}, can be transformed through a conjugation with a
matrix $P\in \mathrm{SO}(4)$ into one of the two inequivalent
(under $\mathrm{SO}(4)$ conjugation) sets of matrices $Y_{\alpha}$
and $\what{Y}_{\alpha}$, $\alpha=1,2,3$. (Note that indeed any
skew-symmetric matrix in $\R^4$ is written as a sum of the
$Y_\alpha$ and of the $\what{Y}_\alpha$.) This corresponds to the
$\mathfrak{su}(2)$ algebra having two irreducible representations
in $\R^4$.

A similar result holds in $\R^{4 n}$: in this case one acts with
$G = \mathrm{SO}(4n)$, and the \HK structures can be transformed
into some direct sum of the above standard ones; we stress in this
sum there will in general be blocks of each orientation. More
precisely we have the following

\medskip\noindent
{\bf Lemma 1.} {\it Given any quaternionic structure $\{
J_{\alpha} \}$ in $\R^{4n}$, there exists a conjugation given by a
regular matrix $P \in \mathrm{SO} (4n)$, such that
$\widetilde{J}_\alpha := P J_{\alpha}P^{-1}$ are diagonal $4\times
4$ block matrices, and the blocks in the diagonal are equal to
either $Y_{\alpha}$ or $\what{Y}_{\alpha}$.}

\medskip\noindent {\bf Proof.} As we are in Euclidean spaces,
$\nabla J_\alpha = 0$ means the $J_\alpha$ are actually constant;
thus they provide a (real, quaternionic) representation of
$\mathrm{SU}(2)$. This can be decomposed as the sum of real
quaternionic irreducible representations, which are well known
(see e.g. chap.8 of \cite{Ki76}) to be four dimensional. See also
the discussion below. \EOP

\subsection{Hyperk\"ahler and strongly \HK maps}

Let us now consider (strongly) \HK maps in Euclidean spaces;
in this case we can be quite more specific, due to the specially
simple metric and the triviality of the Levi-Civita connection.

If $g$ is the matrix associated to the metric and $\Lambda$ is the
Jacobian of a transformation $\varphi$ in $\V$, the change in the
metric is \be g \ \to \ \wt{g} \ = \ \Lambda \, g \ \Lambda^T \ ; \ee
for the Euclidean metric $g = I_4$ and hence $ \wt{g} = \Lambda
\Lambda^T$; thus $\wt{g} = g$ requires
\be  \Lambda \,
\Lambda^T   \equiv \ I_4 \ee
and it should be $\Lambda (x)
\in \mathrm{O}(4n)$ (or $\Lambda (x) \in \mathrm{SO}(4n)$ if we
want to keep the orientation) for any $x\in \V$.

Moreover, as both the original and the transformed complex
structures should be covariantly constant, it should also be
$\nabla \Lambda = 0$: but since here $\nabla$ is the trivial
connection, this means $\partial_i \Lambda (x) = 0$ for all $i =
1,..., 4 n$, hence $\Lambda$ is constant.

This shows at once that \HK and strongly \HK maps
should be constant orthogonal (or special orthogonal if we want to
preserve orientation) ones, in full generality.

\subsection{Strongly \HK maps}

Let us now consider in detail strongly \HK maps, and
represent the $J_\alpha$ by means of the corresponding matrices in
coordinates. Regarding the preservation of the \HK structure, we
should impose in this (strong invariance) case \be\label{strong}
\Lambda \, J_{\alpha} \, \Lambda^{-1} \ = \ J_{\alpha} \ ; \ee
this means that $\Lambda$ should commute with each of the
$J_{\alpha}$. Since these matrices are a representation of
$\mathfrak{su}(2)$, we could apply representation theory to this
problem.

Indeed, let us consider a set of three $m \times m$ real matrices,
$J_{\alpha}$, $\alpha=1,2,3$ satisfying $ J_{\alpha} J_{\beta} =
\epsilon_{\alpha\beta\gamma}J_{\gamma}-\delta_{\alpha\beta}I_m$.
This relation implies $[J_{\alpha} , J_{\beta}] = 2
\epsilon_{\alpha\beta\gamma} J_{\gamma}$, stating that the
matrices $ \Gamma_{\alpha}=(1/2) J_{\alpha}$, $ \alpha=1,2,3 $
form a representation ${\mathcal R}$ of $\mathfrak{su}(2)$.
 But it also yields $ J_{\alpha}^2=-I_m$, $\Gamma_{\alpha}^2=-(1/4)I_m
$. This property implies that the eigenvalues of the Casimir
$-\sum_{\alpha}\Gamma_{\alpha}^2$ are $(3/4)$ and then the
representation is (complex) reducible into a direct sum of a
certain number of $(\frac12)$ representations:
$$ {\mathcal R} \ = \ \left( \frac12 \right) \oplus \cdots \oplus
\left( \frac12 \right) \ . $$
Since we have real matrices, the representation is real reducible
to a diagonal block form, with $4\times 4$ blocks, each of them
associated to a $(\frac12)\oplus (\frac12)$ representation of
$\mathfrak{su}(2)$ with real matrices and the dimension is $m=4n$.

\subsection{Quaternionic maps}

Let us now consider quaternionic (or hyperk\"ahler, see Definition
5) maps. The requirement to map $\Jb$ into a possibly different,
but equivalent, structure $\wt{\Jb}$ implies that \be\label{rest2}
\wt{J}_{\alpha} \ = \ \Lambda J_{\alpha} \Lambda^{-1} \ = \
\sum_{\beta=1}^3 R_{\alpha\beta} J_{\beta}, \quad \sum_{\beta=1}^3
R_{\alpha\beta}^2 = 1 \ , \ \quad \alpha=1,2,3 \ . \ee

In the same way as in the strong version, the first condition
implies that the matrices $\Lambda$ should be in $\mathrm{O}(4n)$,
and in $\mathrm{SO}(4n)$ if we want to leave invariant the
orientation. As for the second condition, it yields the following
constraint. The new matrices $\wt{J}_{\alpha}$ should satisfy the
quaternionic relations \eqref{quatcond}, i.e. \be
\wt{J}_{\alpha}\wt{J}_{\beta} \ = \
\epsilon_{\alpha\beta\gamma}\wt{J}_{\gamma} \ - \
\delta_{\alpha\beta} I \ . \ee Substituting (sum over repeated
indices is assumed) \be R_{\alpha\mu} R_{\beta\nu} J_{\mu}
J_{\nu} = \epsilon_{\mu\nu\rho} R_{\alpha\mu} R_{\beta\nu}
J_{\rho} + \delta_{\mu\nu} R_{\alpha\mu} R_{\beta\nu} I =
\epsilon_{\alpha\beta\gamma} R_{\gamma\rho} J_{\rho} -
\delta_{\alpha\beta} I \ , \ee we obtain \be R_{\alpha\mu}
R_{\beta\mu} = \delta_{\alpha\beta}, \quad \epsilon_{\mu\nu\rho}
R_{\alpha\mu} R_{\beta\nu} = \epsilon_{\alpha\beta\gamma}
R_{\gamma\rho} \ . \ee

The first condition means that the matrix $R$ is an element of
$\mathrm{O}(3)$. The second one means that the vector product of
its first and second column is the third one, which yields
$R\in\mathrm{SO}(3)$. Then, in the end we obtain the equation
 \be\label{finit2}
\Lambda J_{\alpha} \Lambda^{-1}= \sum_{\beta=1}^3 R_{\alpha\beta}
J_{\beta}, \quad \alpha=1,2,3, \ \ \ \Lambda\in\mathrm{SO}(4n),
\quad R \in \mathrm{SO}(3) \ . \ee

Thus our problem is to determine which $\Lambda \in
\mathrm{SO}(4n)$ will satisfy equation \eqref{finit2} for a
certain $R \in \mathrm{SO}(3)$ (which is fixed by $\Lambda$).
The equation \eqref{finit2} will also be called the
{\it finite invariance equation}.

\subsection{The infinitesimal approach\label{sec53}}

It will be convenient, in particular in the high-dimensional case, to approach this problem from the infinitesimal point
of view.

At first order in a certain parameter $\varepsilon$, we have \be
\Lambda=I_{4n}+\varepsilon X,\quad X+X^T=0, \ee and, in terms of
the same parameter, \be R=I_3+\varepsilon \L,\quad \L+\L^T=0 \ . \ee

Equation \eqref{finit2}, for any quaternionic structure
$J_{\alpha}$, is then written at the infinitesimal level as
 \be (I_{4n}+\varepsilon X)
J_{\alpha} (I_{4n}-\varepsilon X)=
\sum_{\beta=1}^3(\delta_{\alpha\beta}+\varepsilon
\L_{\alpha\beta})J_{\beta} \ , \ \ \alpha=1,2,3,
\ee
with $X\in\mathcal{M}_{4n}(\R)$, $\L\in
\mathcal{M}_3$; $X+X^T=0$, $\L+\L^T=0$. That is, at first order in
$\varepsilon$ \be \label{trans} [X,J_{\alpha}]= \sum_{\beta=1}^3
\L_{\alpha\beta}J_{\beta},\quad \alpha=1,2,3. \ee

The equation \eqref{trans} will also be called the
{\it infinitesimal invariance equation}, or shortly
(as we will mainly work in the infinitesimal approach)
the {\it invariance equation}.

The main result of this note is the solution of these equations
for Euclidean spaces, i.e. for $\V=\R^{4n}$ with the Euclidean
metric $g = I_{4n}$.

Note that \eqref{trans} should be seen as an equation for $X$ {\it and} $\L$; on the other hand if we fix $X$, i.e. if we consider a given \HK transformation, we can easily find $\L$ in terms of $X$.

As mentioned above, our main task is to characterize the group of invariance maps for $(\V,g,\Jb)$ when $(\V,g)$ is $\R^{4n}$ with the Euclidean metric. In the infinitesimal approach, we will of course look for the Lie algebra of this group; this will be called the {\it invariance algebra} and denoted as $\mathfrak{L}$.   More specifically, we will denote by $\mathfrak{L}_n$ the invariance algebra for the \HK structures in the Euclidean space $\R^{4n}$.

In order to grasp the problem and the approach to its solution, we find convenient to first consider the simplest (and somehow degenerate) case $n=1$, which we do in the next section; and then the first non-degenerate case $n=2$ in Section \ref{sec:R8}, before tackling the general case in Section \ref{sec:R4n}.

In the following we will systematically use standard cartesian coordinates on the
manifold $\R^{4n}$, and represent the tensors $J_{\alpha}$ by real $4n$-dimensional matrices in the chosen coordinate system without further notice.

\section{The space $\R^4$}
\label{sec:R4}

When the manifold is $\R^4$, the explicit characterization of
quaternionic maps can be obtained in a simple way via either the
infinitesimal approach sketched above, or directly working at the
finite level. The arguments used in the discussion and the proof
below are well known, but keeping them in mind will help in the
study of higher dimensional cases.

\subsection{The infinitesimal approach}

Any skew symmetric $4 \times 4$ matrix $X$ is necessarily a linear
combination of the two sets $Y_{\alpha}$ and $\what{Y}_{\alpha}$,
$\alpha=1,2,3$, given above; we recall these satisfy
$[Y_{\alpha},\what{Y}_{\beta}]=0$. Thus we have
 \be X \ = \ \frac12 \, \sum_{\beta=1}^3 \, c_{\beta} \, Y_{\beta} \ + \
 \frac12 \, \sum_{\beta=1}^3 \, \what{c}_{\beta} \, \what{Y}_{\beta}, \ee
and the invariance equation for the positively oriented
standard structure is
 \be
\bigg[ \frac12 \sum_{\beta=1}^3 c_{\beta} Y_{\beta} + \frac12 \sum_{\beta=1}^3 \what{c}_{\beta} \what{Y}_{\beta} , Y_{\alpha}\bigg] \ = \ \sum_{\beta=1}^3
\L_{\alpha\beta}Y_{\beta} \, , \ \ \ \alpha=1,2,3. \ee
Then, for $\alpha=1,2,3$,
\be
\frac12 \, \sum_{\beta=1}^3 \, c_{\beta} \ [Y_{\beta},Y_{\alpha}
] \ = \ \sum_{\beta=1}^3 \, \L_{\alpha\beta} \, Y_{\beta} \ , \quad
  \sum_{\beta=1}^3 \, c_{\beta} \, \sum_{\gamma=1}^3 \, \epsilon_{\beta\alpha\gamma} \, Y_{\gamma} \ = \ \sum_{\gamma=1}^3 \, \L_{\alpha\gamma} \ Y_{\gamma} \ , \ee
and finally,
\be \label{defM}
\L_{\alpha\beta} \ = \ \sum_{\gamma=1}^3 \, \epsilon_{\alpha\beta\gamma} \, c_{\gamma} \ , \ \ \ \ \alpha,\beta=1,2,3. \ee
It follows from this that we have the

\medskip\noindent
{\bf Theorem 1.} {\it The invariance algebra for any \HK structure
in $(\V,g) = (\R^4,I_4)$ is $\mathfrak{L}_1 = \mathfrak{so}(4)
\simeq \mathfrak{su}(2) \times \mathfrak{su}(2)$.}

\medskip\noindent {\bf Proof.} As seen above, any \HK structure can
be reduced to either the positively or the negatively oriented
standard structure; so it suffices to consider these. We will
consider the positively oriented structure; the discussion for the
negatively oriented one is exactly the same, upon interchanging
the role of the  $Y_\alpha$ and of the $\what{Y}_\alpha$.
 As we have noted above, the effective group, rotating
$Y_{\alpha}$, is a $\mathrm{SO}(3)$ subgroup, with a Lie algebra
generated by the matrices $Y_{\alpha}$. The other $\mathrm{SO}(3)$
subgroup, which is generated by the matrices $\what{Y}_{\alpha}$,
leaves the matrices $J_{\alpha}$ invariant. The result can be
understood in terms of pure group or Lie algebra theory. In fact,
the group $\mathrm{SO}(4)$ is not a simple group but the direct
product of two $\mathrm{SO}(3)$ groups. Its Lie algebra has a real
representation given by $4\times 4$ matrices which splits into the
direct sum of two $\mathfrak{su}(2)$ algebras. Since they commute,
the action of the whole algebra through the adjoint action is
reduced to the action of one of the subalgebras on itself. This is
the reason why $\L$, see equation \eqref{defM}, is in fact in the
3-dimensional representation of $\mathfrak{so}(3) \simeq
\mathfrak{su}(2)$, the action of the other algebra being trivial.
\EOP

\subsection{The finite approach}

In this simple case, we could actually solve the problem using
finite transformations and the real version of Schur lemma (see
e.g. \cite{Ki76}, chapter 8). From a practical point of view, we
can directly compute the matrices commuting with this
representation. The solution of equation \eqref{strong} is  \be
\Lambda=
\begin{pmatrix}
 a & -d & b & -c \\
 d & a & c & b \\
 -b & -c & a & d \\
 c & -b & -d & a
\end{pmatrix}  =aI_4
+b \begin{pmatrix}
 0 & \sigma_0 \\
 -\sigma_0 & 0
\end{pmatrix}
+c \begin{pmatrix}
 0 & -\ri\sigma_2 \\
 -\ri\sigma_2 & 0
\end{pmatrix}
+d \begin{pmatrix}
 -\ri \sigma_2 & 0 \\
 0 & \ri\sigma_2
\end{pmatrix}  \ . 
\ee Here the $\sigma_i$ are the standard (complex, two dimensional) Pauli
matrices \cite{LL}; in particular
$$ \sigma_0 \ = \ \begin{pmatrix} 1 & 0 \\ 0 & 1 \end{pmatrix} \ , \ \ 
\sigma_2 \ = \ \begin{pmatrix} 0 & - i \\ + i & 0 \end{pmatrix} \ . $$

This real matrix (with $\det \Lambda=(a^2 + b^2 + c^2 + d^2)^2$)
must be in $\mathrm{O}(4)$, and (as $a,b,c,d$ are real) the only
condition to be satisfied is \be a^2+b^2+c^2+d^2=1, \ee and then,
it is in $\mathrm{SO}(4)$. The matrices $\Lambda$ form a subgroup
of $\mathrm{SO}(4)$, in fact, the subgroup generated by the
negatively oriented standard form of the quaternionic structure.
The coefficients $a,b,c,d$ could depend on the point $x\in \R^4$.
Then, the set of matrices leaving invariant the positively
oriented standard quaternionic structure is, in each point of
$\R^4$, the quaternion group (the set of quaternion units)
generated by the negatively oriented standard quaternionic
structure, which is a subgroup of $\mathrm{SO}(4)$, in fact
$\mathrm{SO}(3)$.

\medskip\noindent {\bf Remark 6.} It may be useful to show
explicitly (in a simple case) the interrelation between the
matrices $R$ and $\Lambda$, also to illustrate the situation in
the finite approach. Every element $\Lambda\in \mathrm{SO}(4)$ can
be written in a $2\times 2$ block-diagonal form (after a
conjugation with an appropriate $P\in\mathrm{SO}(4)$). Let us
consider those elements (pure rotations in the 2-dimensional
planes $(x_1,x_2)$ and $(x_3,x_4$)) of the form
 \be
\Lambda_0= \begin{pmatrix} \cos\phi_1 & -\sin\phi_1 & 0 &
0 \\ \sin \phi_1 & \cos\phi_1 & 0 &0 \\ 0 & 0 & \cos \phi_2 &
-\sin \phi_2 \\ 0 & 0 & \sin\phi_2 & \cos \phi_2
\end{pmatrix}  ,\quad \phi_1,\phi_2\in \R,\qquad \Lambda=P\Lambda_0P^{-1}.
\ee

If in this basis the quaternionic structure is the positively
oriented standard one, the matrices $Y_{\alpha}$ are transformed
under this matrix $\Lambda_0$ as
\begin{align}
Y_1&\to Y_1 \\
Y_2&\to Y_2\cos(\phi_1+\phi_2)-Y_3\sin(\phi_1+\phi_2) \\
Y_3&\to Y_2\sin(\phi_1+\phi_2)+Y_3\cos(\phi_1+\phi_2),
\end{align}
that is, the matrix $R$ is
\be R \ = \
 \begin{pmatrix}
 1 & 0 & 0 \\
 0 & \cos(\phi_1 +\phi_2) & -\sin (\phi_1  +\phi_2) \\
 0 & \sin(\phi_1 +\phi_2) & \ \cos(\phi_1 +\phi_2)
\end{pmatrix}  .
\ee
The generic situation will be
$$
\Lambda Y_{\alpha}\Lambda^{-1} \ = \ P (\Lambda_0 ( P^{-1} Y_{\alpha} P) \Lambda_0^{-1}) P^{-1} \
= \ \sum_{\beta=1}^3 R_{\alpha \beta} Y_{\beta} \ ,
$$
$$\Lambda_0(P^{-1}Y_{\alpha}P)\Lambda_0^{-1}
=\sum_{\beta=1}^3R_{\alpha\beta}P^{-1}Y_{\beta}P \ ;
$$
or,
\be
\Lambda_0 J_{\alpha} \Lambda_0^{-1}
= \sum_{\beta=1}^3 R_{\alpha\beta} J_{\beta},\quad J_{\alpha}=P^{-1}Y_{\alpha}P ;
\ee
it is not easy to write  $R$ in terms of $\Lambda$ in an explicit way.

Finally, note that the set $\{I_4,Y_1,Y_2,Y_3\}$ provides a real
representation of the quaternion units $1,\ri,\rj,\rk$. An
analysis using quaternion algebra would  yield the same results we
have got above.\EOR


\section{The space $\R^8$}
\label{sec:R8}

In the previous Section we discussed the case $n=1$, which is
somewhat degenerate in that the \HK structures in standard form
consist of a single block. In this section we will tackle the
first non-degenerate case, i.e. $n=2$ or $\R^8$; this will present
the difficulties met in the general one $\R^{4n}$, but for it the
identification of the invariance algebra $\mathfrak{L}$ is still
rather straightforward.

First of all, we note that albeit Lemma 1 would lead us to deal
with four different types of \HK structures, the invariance groups
(and algebras) for them are isomorphic; this follows from
different classes of standard \HK structures being conjugated by
the action of matrices in $\mathrm{O}(8)$.

\medskip\noindent
{\bf Lemma 2.} {\it All \HK structures in Euclidean $\R^8$ are
conjugated under $\mathrm{O}(8)$.}

\medskip\noindent {\bf Proof.} Using Lemma 1, the quaternionic
structure $J_{\alpha}$ (which we recall is necessarily constant)
in $\R^8$ endowed with the Euclidean metric, can be reduced to one
of the following types (in the third case the order of the blocks
can be reversed): \be Y^{(1)}_{\alpha}= \begin{pmatrix} Y_{\alpha}
\\ & Y_{\alpha}\end{pmatrix}  ,\quad Y^{(2)}_{\alpha}=
\begin{pmatrix} \what{Y}_{\alpha} \\ &
\what{Y}_{\alpha}\end{pmatrix}  ,\quad Y^{(3)}_{\alpha}=
\begin{pmatrix} Y_{\alpha} \\ & \what{Y}_{\alpha}\end{pmatrix}  .
\ee

In fact there exist a four dimensional matrix $Q$ which satisfies
$Q Q^T=\lambda I_4$ and can be chosen in $\mathrm{O}(4) \backslash
\mathrm{SO} (4)$, i.e. in the elements of $\mathrm{O}(4)$ with
determinant equal to $-1$, such that $Y_{\alpha} = Q^{-1}
\what{Y}_{\alpha} Q$ for $\alpha=1,2,3$; here $Q^{-1}=Q^T$, $\det
Q=-1$. Then, if $Q_2=\mathrm{diag}(Q,Q) \in \mathrm{O}(8)$ and
$Q_3=\mathrm{diag}(I_4,Q) \in \mathrm{O}(8)$, we get \be
Q_2^{-1}Y^{(2)}_{\alpha}Q_2=Y^{(1)}_{\alpha},\quad
Q_3^{-1}Y^{(3)}_{\alpha}Q_3=Y^{(1)}_{\alpha}. \ee This shows that
all the standard \quaternionic structures -- and hence, in view of
Lemma 1, all the \quaternionic structures -- in $\R^8$ are
conjugated, as stated. \EOP

\medskip\noindent
{\bf Theorem 2.} {\it For any \HK structure in $(\V,g) =
(\R^8,I_8)$, the invariance algebra is $\mathfrak{L}_2 =
\mathfrak{su}(2) \otimes \mathfrak{sp}(2)$.}

\medskip\noindent {\bf Proof.} In view of Lemma 2, we can just deal
with $Y_\alpha^{(1)}$. An orthogonal transformation, leaving
invariant the Euclidean metric $I_8$ is an element of
$\mathrm{O}(8)$, satisfying \be \Lambda \Lambda^T=I_8,\quad
\det\Lambda=1; \ee at the infinitesimal level the invariance of
the quaternionic relations imposes, as above, \be [X,J_{\alpha}]=
\sum_{\beta=1}^3 \L_{\alpha\beta}J_{\beta},\quad \alpha=1,2,3, \ee
where $X = - X^T \in \mathcal{M}_8$ and $\L = - \L^T \in
\mathcal{M}_3$; note that here $\Lambda \in {\mathrm O}(8)$
implies actually $X \in \mathfrak{so}(8)$.

As in the previous case, the three matrices $J_{\alpha}$ generate
an $\mathfrak{su}(2)$ algebra which is contained in the algebra
$\mathfrak{so}(8)$ (of dimension 28) of $\mathrm{SO}(8)$. However,
in this case, the situation is not so simple, because the whole
$\mathfrak{so}(8)$ cannot be generated by the quaternionic
matrices (even considering both orientations and their
combinations in the $8\times 8$ matrices). A basis of
$\mathfrak{so}(8)$ is: \be
 \begin{pmatrix} Y_{\alpha} & 0 \\ 0& 0
\end{pmatrix}  ,\;  \begin{pmatrix} 0 & 0 \\ 0&  Y_{\alpha}
\end{pmatrix}  ,\;  \begin{pmatrix}  \what{Y}_{\alpha} & 0 \\ 0& 0
\end{pmatrix}  ,\;  \begin{pmatrix} 0 & 0 \\ 0& \what{Y}_{\alpha}
\end{pmatrix}  ,\;
 \begin{pmatrix} 0 & Y_{\alpha} \\ Y_{\alpha}& 0
\end{pmatrix}  ,\;  \begin{pmatrix} 0 & \what{Y}_{\alpha} \\ \what{Y}_{\alpha}& 0
\end{pmatrix}  ,\;  \begin{pmatrix} 0 & S_i \\ -S_i& 0
\end{pmatrix}  .
\ee with $\alpha=1,2,3$, and $S_i$, $i=1,\ldots,10$, the set of
$4\times 4$ elementary symmetric matrices (that is, $E_{jj}$ and
$E_{jk}+E_{kj}$, where $E_{jk}$ is the elementary matrix with 1 in
the position $jk$ and $0$ elsewhere).

Let us first consider the structure $Y^{(1)}_{\alpha}$ and the
equation \eqref{trans}. If we write \be X= \begin{pmatrix} A & B
\\ -B^T& C
\end{pmatrix}  ,\quad A+A^T=0,\quad C+C^T=0
\ee
we get
\be
\left[ \begin{pmatrix} A & B \\ -B^T& C
\end{pmatrix}  , \begin{pmatrix} Y_{\alpha} & 0 \\ 0& Y_{\alpha}
\end{pmatrix}  \right]=\sum_{\beta=1}^3 \L_{\alpha\beta} \begin{pmatrix} Y_{\beta} & 0 \\ 0& Y_{\beta}
\end{pmatrix}
\ee and the relations \be [A,Y_{\alpha}]=\sum_{\beta=1}^3
\L_{\alpha\beta}Y_{\beta},\quad [C,Y_{\alpha}]=\sum_{\beta=1}^3
\L_{\alpha\beta}Y_{\beta},\quad [B,Y_{\alpha}]=0. \ee Using the
previous results in dimension 4, we obtain, from the first
relation \be
A=\frac12\sum_{\beta=1}^3a_{\beta}Y_{\beta}+\frac12\sum_{\beta=1}^3\what{a}_{\beta}\what{Y}_{\beta}
,\quad
C=\frac12\sum_{\beta=1}^3c_{\beta}Y_{\beta}+\frac12\sum_{\beta=1}^3\what{c}_{\beta}\what{Y}_{\beta}
,\quad
\L_{\alpha\beta}=\sum_{\gamma=1}^3\epsilon_{\alpha\beta\gamma}a_{\gamma},
\ee and then $a_{\beta}=c_{\beta}$, while $\what{a}_{\beta}$ and
$\what{c}_{\beta}$ are arbitrary constants.

As for $B$, we get (here $B_S$ is the symmetric part of $B$)
\be
B=\frac12\sum_{\beta=1}^3b_{\beta}Y_{\beta}+\frac12\sum_{\beta=1}^3\what{b}_{\beta}\what{Y}_{\beta}+B_S,
\quad [B,Y_{\alpha}]=0 \;\Rightarrow\; b_{\alpha}=0,\quad
B_S=\lambda I_4, \ee and $\what{b}_{\beta}$ are arbitrary
constants.

These results provide a subalgebra of $\mathfrak{so}(8)$ with basis
\be\label{basis}
 \begin{pmatrix} Y_{\alpha} & 0 \\ 0& Y_{\alpha}\end{pmatrix}  ,\quad
 \begin{pmatrix} \what{Y}_{\alpha} & 0 \\ 0& 0\end{pmatrix}  ,\quad
 \begin{pmatrix} 0 & 0 \\ 0& \what{Y}_{\alpha}\end{pmatrix}  ,\quad
 \begin{pmatrix} 0 & \what{Y}_{\alpha} \\ \what{Y}_{\alpha}& 0\end{pmatrix}  ,\quad
 \begin{pmatrix} 0 & I_4 \\ -I_4& 0\end{pmatrix}  .
\ee

In fact, the matrix $\L$ is different from zero only for the
generators
\be\label{rot}  \begin{pmatrix} Y_{\alpha} & 0
\\ 0& Y_{\alpha}\end{pmatrix}  , \ee
and the corresponding
algebra is $\mathfrak{su}(2)$. The other matrices,  which generate
a Lie algebra of dimension 10 which commutes with the algebra
$\mathfrak{su}(2)$ generated by the matrices \eqref{rot}, leave
invariant each of the matrices $J_{\alpha}$, $\alpha=1,2,3$. The
commutation table appears in Appendix \ref{app1}, see Table \ref{algC2}.

It is an easy task to construct the adjoint representation and the
Killing form and this allows to identify the algebra as a real
compact semisimple Lie algebra, $\mathfrak{sp}(2)$, whose complex
extension is isomorphic to the Lie algebra $C_2$ (or $B_2$) in the
Cartan classification. The details of the computations are
contained in Appendix \ref{app1}.

If we consider the structure $Y^{(i)}_{\alpha}$, $i=2,3$, the
equation to be solved is \be [X^{(i)},Y^{(i)}_{\alpha}]=
\sum_{\beta=1}^3 \L_{\alpha\beta}Y^{(i)}_{\beta},\quad
\alpha=1,2,3,\; i=2,3, \ee and then \be
[X^{(i)},Q_iY^{(1)}_{\alpha}Q_i^{-1}]= \sum_{\beta=1}^3
\L_{\alpha\beta}Q_iY^{(1)}_{\beta}Q_i^{-1} \ee or \be
[Q_i^{-1}X^{(i)}Q_i,Y^{(1)}_{\alpha}]= \sum_{\beta=1}^3
\L_{\alpha\beta}Y^{(1)}_{\beta} , \ee and the algebra formed by
the matrices $X^{(i)}$ is now conjugated to the one we get for the
first structure  $Y^{(1)}_{\alpha}$. In fact, in the case
$Y_{\alpha}^{(2)}$, the roles of $Y_{\alpha}$ and
$\what{Y}_{\alpha}$ are simply exchanged. The basis for the
subalgebra  is \be  \begin{pmatrix} \what{Y}_{\alpha} &
0 \\ 0& \what{Y}_{\alpha}\end{pmatrix}  ,\quad
 \begin{pmatrix} Y_{\alpha} & 0 \\ 0&
0\end{pmatrix}  ,\quad  \begin{pmatrix} 0 & 0 \\ 0&
Y_{\alpha}\end{pmatrix}  ,\quad  \begin{pmatrix} 0 &
Y_{\alpha} \\ Y_{\alpha}& 0\end{pmatrix}  ,\quad
 \begin{pmatrix} 0 & I_4 \\ -I_4& 0\end{pmatrix}  , \ee
and the invariance algebra is again $\mathfrak{sp}(2)$.

Finally, for the third possible structure  $Y_{\alpha}^{(3)}$, we
also get $\mathfrak{sp}(2)$ and a basis is: \be
 \begin{pmatrix} Y_{\alpha} & 0 \\ 0&
\what{Y}_{\alpha}\end{pmatrix}  ,\quad  \begin{pmatrix}
\what{Y}_{\alpha} & 0 \\ 0& 0\end{pmatrix}  ,\quad
 \begin{pmatrix} 0 & 0 \\ 0&
Y_{\alpha}\end{pmatrix}  ,\quad  \begin{pmatrix} 0 & Z_i
\\ - Z^T_i& 0\end{pmatrix}   , \ee where $Z_i$, $i=1,\ldots,4$
are the matrices {\small\be  \begin{pmatrix}
 1 & 0 & 0 & 0 \\
 0 & 0 & 1 & 0 \\
 0 & -1 & 0 & 0 \\
 0 & 0 & 0 & -1 \\
\end{pmatrix}
,\;
 \begin{pmatrix}
 0 & 1 & 0 & 0 \\
 0 & 0 & 0 & 1 \\
 1 & 0 & 0 & 0 \\
 0 & 0 & 1 & 0 \\
\end{pmatrix}
,\;
 \begin{pmatrix}
 0 & 0 & 1 & 0 \\
 -1 & 0 & 0 & 0 \\
 0 & 0 & 0 & 1 \\
 0 & -1 & 0 & 0 \\
\end{pmatrix},
\;
\begin{pmatrix}
 0 & 0 & 0 & -1 \\
 0 & 1 & 0 & 0 \\
 0 & 0 & 1 & 0 \\
 -1 & 0 & 0 & 0 \\
\end{pmatrix},
\ee} \noindent which correspond to matrices $Q$ intertwining the
two sets $Y_{\alpha}$ and $\what{Y}_{\alpha}$. The invariance
algebra is still the same. This concludes the proof. \EOP

\medskip\noindent {\bf Remark 7.} We have obtained above
$\mathfrak{L}_1 =
\mathfrak{so}(4)=\mathfrak{su}(2)\oplus\mathfrak{su}(2)$. The
first $\mathfrak{su}(2)$ corresponds to the strong invariance
algebra and the second one to the regular invariance algebra. In
the $\R^8$ case, as we have seen, the invariance algebra is
$\mathfrak{L}_2 = \mathfrak{su}(2)\oplus\mathfrak{sp}(2)$. In
fact, the case $\R^4$ has exactly the same structure, since
$\mathfrak{sp}(1)\approx\mathfrak{su}(2)$. \EOR


\section{The general case: $\R^{4n}$}
\label{sec:R4n}

We are now ready to tackle the general case, i.e. the Euclidean
space $\R^{4n}$. We will face two difficulties: the notation,
which is necessarily rather cumbersome, and more substantially the
identification of the invariance algebra $\mathfrak{L}$. In our
discussion we will again rely on Lemma 1, and hence consider \HK
structures in standard form; and will make use of the notation
introduced in the discussion of the $\R^8$ case.

Given two square matrices $A$ and $B$, respectively of dimension
$m$ and $n$, we choose the basis of their tensorial product in
such a way that the $m \times m$ block matrix (with $n \times n$
blocks) is \be A \otimes B \ = \ (a_{ij}B) \ , \ \ A=(a_{ij}) \ ,
\ee and denote by $E_{ij}$ the usual elementary matrix: \be
(E_{ij})_{kl} \ := \ \delta_{ik} \delta_{jl} \ , \ \ E_{ij} E_{kl}
\ = \ \delta_{jk} E_{il} \ . \ee Then $E_{ij}\otimes B$ is a $nm
\times nm$ matrix formed by $n \times n$ blocks,  where the $ij$
block is equal to $B$, and all elements elsewhere are zero.

We recall that the invariance condition for a quaternionic
structure in $\R^{4n}$ can be expressed by \eqref{trans}; recall
also that $X = - X^T \in \mathcal{M}_{4n}$, $\L = - \L^T  \in
\mathcal{M}_3$.

Using Lemma 1, any quaternionic structure in $\mathbf{R}^{4n}$ is
conjugated to \be J_{\alpha}=\sum_{i=1}^n E_{ii}\otimes
J^{(i)}_{\alpha}, \quad \alpha=1,2,3, \ee where $J^{(i)}_{\alpha}$
is any of the two nonequivalent quaternionic structures in
$\mathbf{R}^4$, i.e. either $Y_{\alpha}$ or $\what{Y}_{\alpha}$,
and we can reduce our problem to the simplest case of
block-diagonal quaternionic structures.

However, we can have in the diagonal both kinds of orientations, a
difficulty which can be easily surmounted  using the fact, already
used in the case $\R^8$, that  there exist an orthogonal matrix
$Q$, with $\det Q=-1$, such that
$Y_{\alpha}=Q^{-1}\what{Y}_{\alpha}Q$, for $\alpha=1,2,3$; see
Lemma 2 above. We actually extend this to a higher dimensional
setting; the proof of this is straightforward and hence omitted.

\medskip\noindent
{\bf Lemma 3.} {\it Let $J_{\alpha}$ be a quaternionic structure
in $\R^{4n}$ set in a $4\times 4$ block-diagonal form,
$J_{\alpha}=\sum_{i=1}^n E_{ii}\otimes J^{(i)}_{\alpha}$. Then,
the block-diagonal matrix \be
\mathbf{Q}=\mathrm{diag}(Q_1,Q_2,\ldots,Q_n) \ , \ \
Q_i=\begin{cases}
I_4 & \hbox{if} \; J_{\alpha}^{(i)}=Y_{\alpha}\, ,  \\
Q & \hbox{if}\;  J_{\alpha}^{(i)}=\what{Y}_{\alpha}\, , \end{cases}
\ee
satisfies
\be
\mathbf{Q}^{-1}J_{\alpha}\mathbf{Q}=\sum_{i=1}^n E_{ii}\otimes Y_{\alpha}.
\ee}

We need another preliminary result before going into the explicit
computation of the invariance algebra, generalizing Lemma 2 and
Lemma 3; this will allow us to reduce all the different
orientation cases to the positively oriented one.

\medskip\noindent
{\bf Lemma 4.} {\it The invariance algebras of all the
quaternionic structures in $\R^{4n}$ are isomorphic.}

\medskip\noindent {\bf Proof.} We follow, with obvious
modifications, the argument used in the case $\R^8$. Thanks to
Lemma 1, we can just consider structures in standard form. The
invariance equation is \be [X,J_{\alpha}]= \sum_{\beta=1}^3
\L_{\alpha\beta}J_{\beta},\quad \alpha=1,2,3. \ee If we conjugate
the quaternionic structure, $\wt{J}_{\alpha} = U J_{\alpha}
U^{-1}$, we get \be [X,U^{-1}\wt{J}_{\alpha}U]= \sum_{\beta=1}^3
\L_{\alpha\beta}U^{-1}\wt{J}_{\beta}U,\quad \alpha=1,2,3; \ee that
is, \be [UXU^{-1},\wt{J}_{\alpha}]= \sum_{\beta=1}^3
\L_{\alpha\beta}\wt{J}_{\beta} . \ee Then the algebra generated by
the matrices $X$ is isomorphic via a conjugation to the algebra
generated by $\wt{X}=UXU^{-1}$.

The two operations we make to pass from any quaternionic structure
to the positively oriented block-diagonal one, via a preliminary
reduction to a block-diagonal one with arbitrary orientation, are
conjugations; thus the statement is proved. \EOP

We are now ready to identify the structure of the invariance
algebra $\mathfrak{L}_n$; this is the main result of the present
work; we will split its proof into several lemmas.

\medskip\noindent
{\bf Theorem 3.} {\it The invariance algebra for $(\R^{4n},I_{4n},\Jb)$ is
$\mathfrak{L}_n = \mathfrak{su}(2) \oplus \mathfrak{sp} (n)$.}

\medskip\noindent
{\bf Proof.} We note preliminarily that in order to preserve the
Euclidean metric in $\R^{4n}$, necessarily the infinitesimal
transformation $X = - X^T \in \mathcal{M}_{4n}$ belongs to
$\mathfrak{so}(4n)$. Moreover, Lemma 4 guarantees the invariance
algebra for different \HK structures on $(\V,g) =
(\R^{4n},I_{4n})$ are isomorphic; thus we only have to study the
positively oriented standard one.

That is, we consider the \HK structure given by $\{J_1,J_2,J_3\}$ with
\be
J_{\alpha}=\sum_{i=1}^n E_{ii}\otimes Y_{\alpha}, \quad \alpha=1,2,3.
\ee

In this case, the infinitesimal invariance condition is expressed
by equation \eqref{trans} (where $\L = - \L^T \in \mathcal{M}_3$
is the infinitesimal transformation corresponding to a rotation in
$\R^3$), see Section \ref{sec53} above.

\medskip\noindent
{\bf Lemma 5.} {\it The subalgebra of $\mathfrak{L}_n \subset
\mathfrak{so}(4n)$ of the $X$ satisfying the infinitesimal
invariance equation \eqref{trans} (and hence leaving invariant the
quaternionic structure) has dimension $n(2n+1)+3$, and a basis of
it is provided by

\begin{align*}
& \begin{pmatrix}
Y_{\alpha} \\ & Y_{\alpha} \\  &&\ddots \\ &&& Y_{\alpha}
\end{pmatrix},\quad
\begin{pmatrix}
\what{Y}_{\alpha} \\ & 0 \\  &&\ddots \\ &&& 0
\end{pmatrix},
\ldots,
\begin{pmatrix}
0 \\ & 0 \\  &&\ddots \\ &&& \what{Y}_{\alpha}
\end{pmatrix},
\\ &
 \begin{pmatrix}
0 & \what{Y}_{\alpha} & 0 \\ \what{Y}_{\alpha} & 0& 0 \\ 0 & 0 & 0 \\ & &&\ddots \\ &&&&0
\end{pmatrix},\quad
\begin{pmatrix}
0 & 0& \what{Y}_{\alpha} \\  0 & 0 & 0\\ \what{Y}_{\alpha} & 0 & 0 \\  &&&\ddots \\ &&&&0
\end{pmatrix},\ldots,
\begin{pmatrix}
0\\ &\ddots \\ && 0 & 0 & 0\\&& 0 & 0 & \what{Y}_{\alpha} \\ &&  0 & \what{Y}_{\alpha} & 0
\end{pmatrix},
\\
&\begin{pmatrix}
0 & I_4 & 0 \\ -I_4 & 0& 0 \\ 0 & 0 & 0 \\ & & &\ddots \\ &&&&0
\end{pmatrix},\quad
\begin{pmatrix}
0 & 0 & I_4 \\  0 & 0 & 0\\ -I_4 & 0 & 0 \\  &&&\ddots \\ &&&&0
\end{pmatrix},\ldots,\begin{pmatrix}
0\\ &\ddots \\ && 0 & 0 & 0\\&& 0 & 0 & I_4 \\ &&  0 & -I_4 & 0
\end{pmatrix}.
\end{align*}}

\medskip\noindent {\bf Proof.} The equation \eqref{trans} can be read
in the following way. The matrices $J_{\alpha}$ generate a
$\mathfrak{so}(3)$ algebra. The matrices $X$ obviously form also a
subalgebra $\mathfrak{L}$ of $\mathfrak{so}(4n)$, and
\begin{align}
[[X,\wt{X}],J_{\alpha}]=&
[[X,J_{\alpha}],\wt{X}]-[[\wt{X},J_{\alpha}],X] =
\sum_{\beta}\L_{\alpha\beta}[J_{\beta},\wt{X}] -
\sum_{\beta}\wt{\L}_{\alpha\beta}[J_{\beta},X]\nonumber\\
= & \sum_{\gamma}[\wt{\L},\L]_{\alpha\gamma}J_{\gamma}\, .
\end{align}
Note that equation \eqref{trans} simply states the fact that the
subalgebra $\mathfrak{su}(2)$ generated by the $J_\alpha$ is an
ideal of the  algebra $\mathfrak{L}$.

We can construct a basis of $\mathfrak{so}(4n)$, which has
dimension $2n(4n-1)$, using the matrices $Y_{\alpha}$,
$\what{Y}_{\alpha}$, their products and the tensor products with
the matrices $E_{ij}$.

More explicitly, such a basis is provided by the matrices
\begin{align*}
A_{ij\alpha}=&\frac12(E_{ij}+E_{ji})\otimes Y_{\alpha},\quad
i,j=1,\ldots,n,\quad \alpha=1,2,3
\\
\what{A}_{ij\alpha}=&\frac12(E_{ij}+E_{ji})\otimes
\what{Y}_{\alpha},\quad i,j=1,\ldots,n,\quad \alpha=1,2,3
\\
B_{ij\alpha\beta}=&\frac12(E_{ij}-E_{ji})\otimes
Y_{\alpha}\what{Y}_{\beta},\quad i<j=1,\ldots,n,\quad
\alpha,\beta=1,2,3
\\
C_{ij}=&\frac12(E_{ij}-E_{ji})\otimes I_4,\quad i<j=1,\ldots,n.
\end{align*}
It will be notationally convenient to use unconstrained indices
$i,j$ with the conventions
$$
B_{ij\alpha\beta}=-B_{ji\alpha\beta},\quad  C_{ij}=-C_{ji}, \quad
i>j, \quad B_{ii\alpha\beta}=0,\quad C_{ii}=0.
$$
Some of the commutation relations among these elements will be
used in the sequel and can be computed explicitly:
\begin{align}
[A_{ij\alpha},A_{kk\gamma}]=
&\sum_{\nu}\epsilon_{\alpha\gamma\nu}(\delta_{jk}+\delta_{ik})A_{ij\nu}-
\delta_{\alpha\gamma} (\delta_{jk} -\delta_{ik})C_{ij}
\nonumber \\
[\what{A}_{ij\alpha},A_{kk\gamma}]= &(\delta_{jk}-\delta_{ik})B_{ij\gamma\alpha}
\nonumber \\
[B_{ij\alpha\beta},A_{kk\gamma}]= &
\sum_{\nu}\epsilon_{\alpha\gamma\nu}(\delta_{jk}+\delta_{ik})B_{ij\nu\beta}
- \delta_{\alpha\gamma}(\delta_{jk}-\delta_{ik})\what{A}_{ij\beta}
\nonumber \\
[C_{ij},A_{kk\gamma}]= &(\delta_{jk}-\delta_{ik})A_{ij\gamma}\, . \label{basis4n}
\end{align}

Note that the matrices in the positively oriented quaternionic
structure are in this notation written as \be J_{\alpha} =
\sum_{i} A_{ii\alpha} = \sum_{i} E_{ii} \otimes Y_{\alpha}\ . \ee
With the convention that indices in the coefficients are also
unconstrained, and $b_{ij\alpha\beta}=-b_{ji\alpha\beta}$,
$c_{ij}=-c_{ji}$, the invariance condition for this structure is
thus written in terms of the basis \eqref{basis4n} as
\begin{align}
& \sum_{i,j,k,\alpha}a_{ij\alpha}[ A_{ij\alpha},   A_{kk\gamma}
]+\sum_{i,j,k,\alpha}\what{a}_{ij\alpha}[\what{A}_{ij\alpha},
A_{kk\gamma} ]+\sum_{i,j,k,\alpha,\beta}b_{ij\alpha\beta}[
B_{ij\alpha\beta},  A_{kk\gamma}]+\sum_{i,j,k}c_{ij}[C_{ij},
A_{kk\gamma}]\nonumber \\  & =\sum_{\nu,i}\L_{\gamma\nu}
A_{ii\nu},\quad \gamma=1,2,3.
\end{align}

Substituting in this the commutators computed above (and
understanding all equations are for $\gamma=1,2,3$) we get
\begin{align*}
& \sum_{i,j,k,\alpha}a_{ij\alpha,\nu}
\epsilon_{\alpha\gamma\nu}(\delta_{jk}+\delta_{ik})A_{ij\nu}-\sum_{i,j,k,\alpha
}a_{ij\alpha} \delta_{\alpha\gamma} (\delta_{jk}
-\delta_{ik})C_{ij} +\sum_{i,j,k,\alpha}\what{a}_{ij\alpha}
(\delta_{jk}-\delta_{ik})B_{ij\gamma\alpha} \nonumber \\ & +
\sum_{i,j,k,\alpha,\beta,\nu}b_{ij\alpha\beta}
\epsilon_{\alpha\gamma\nu}(\delta_{jk}+\delta_{ik})B_{ij\nu\beta}-
\sum_{i,j,k,\alpha,\beta}b_{ij\alpha\beta}\delta_{\alpha\gamma}(\delta_{jk}-
\delta_{ik})\what{A}_{ij\beta}
\nonumber \\  &+\sum_{i,j,k}c_{ij}
(\delta_{jk}-\delta_{ik})A_{ij\gamma} =\sum_{\nu,i}\L_{\gamma\nu}
A_{ii\nu};
\end{align*}
upon standard simplification this reduces to \be 2\sum_{i,j
,\alpha,\nu} \epsilon_{\alpha\gamma\nu}  a_{ij\alpha }A_{ij\nu}  +
2 \sum_{i,j ,\alpha,\beta,\nu} \epsilon_{\alpha\gamma\nu}
b_{ij\alpha\beta} B_{ij\nu\beta}    =\sum_{\nu,i}\L_{\gamma\nu}
A_{ii\nu} \ . \ee This should be seen as a matrix equation, i.e. a
set of scalar equations, for the coefficients $a,b$ (note the
$c_{ij}$ cancelled out) and for the matrix elements $\L_{ij}$. As
for the $a_{ijk}$ and the $\L_{ij}$, the solution is
\begin{align}
&a_{ij\alpha}=0,\quad i\neq j
\\
& \L_{12}=2a_{ii3 },\quad  \L_{13} =- 2a_{ii2 } ,\quad  \L_{23}=2a_{ii1}\, .
\end{align}
On the other hand the equations for $b_{ij\alpha\beta}$ have the unique solution
\be
b_{ij\alpha\beta}=0,\quad i,j=1,\dots,n,\; \alpha,\beta=1,2,3.
\ee

The invariance algebra is then formed by the elements of the form
\begin{align}
X =& \sum_{i,j,\alpha}a_{ij\alpha}
A_{ij\alpha}+\sum_{i,j,\alpha}\what{a}_{ij\alpha}\what{A}_{ij\alpha}
+\sum_{i,j}c_{ij}C_{ij} \nonumber
\\
 =& \frac12 \L_{23}J_{1}+\frac12 \L_{31} J_{2}+\frac12 \L_{12} J_{3} \label{X4n}  \\
  &  +\frac12\sum_{i,j,\alpha}\what{a}_{ij\alpha}(E_{ij}+E_{ji})\otimes\what{Y}_{\alpha}+
  \frac12\sum_{i,j}c_{ij}(E_{ij}-E_{ji})\otimes I_4 \ . \nonumber
\end{align}
One can check explicitly that these elements satisfy the
invariance equation.

It seems at first sight that this would leave open the possibility
that other elements are also in the invariance algebra. But
actually the algebra spanned by the elements thus identified is a
maximal subalgebra of $\mathfrak{so}(4n)$; given that obviously
not all elements of $\mathfrak{so}(4n)$ preserve the quaternionic
structure, one is guaranteed to have indeed identified the full
invariance algebra. This concludes the proof of Lemma 5. \EOP

\medskip\noindent
{\bf Theorem 4.} {\it The invariance algebra $\mathfrak{L}_n$ is
the direct sum of two mutually commuting subalgebras, \be
\mathfrak{L}=\mathfrak{su}(2)\oplus \mathfrak{g}, \ee one of them
being the $\mathfrak{su}(2)$ algebra generated by the $J_\alpha$,
and the other being a Lie algebra of dimension $n(2n+1)$.}

\medskip\noindent {\bf Proof.} Obviously there is a subalgebra
generated by the three first diagonal elements in \eqref{X4n},
i.e. generated by
$X_\alpha=\mathrm{diag}(Y_{\alpha},\ldots,Y_{\alpha})$; this is
precisely the $\mathfrak{su}(2)$ subalgebra. (This fact
corresponds to the one, already remarked, that \eqref{trans} means
that the subalgebra generated by the $J_\alpha$ is an ideal in
$\mathfrak{L}$.) It is obvious from \eqref{X4n} that all other
elements also form a subalgebra, and that the two subalgebras
commute due to $[Y_{\alpha},\what{Y}_{\beta}]=0$. The statement on
the dimension of $\mathfrak{g}$ follows by direct inspection. \EOP

\medskip\noindent {\bf Remark 8.} It also follows easily from
$[Y_{\alpha},\what{Y}_{\beta}]=0$ that $\mathfrak{g}$ is actually
the strong invariance algebra for $\Jb$. \EOR
\bigskip

We are left with the task of identifying the Lie algebra
$\mathfrak{g}$; this is not immediate and will require some Lie
algebra theory. We actually know that $\mathfrak{g}$  is equal to
$\mathfrak{sp}(n)$ when $n=1,2$, see Theorem 1 and Theorem 2 (and
Remark 7). It turns out that this is always the case.

\medskip\noindent
{\bf Theorem 5.} {\it The strong invariance algebra for the
standard positively oriented \HK structure on Euclidean $\R^{4n}$
is $\mathfrak{g} = \mathfrak{sp}(n)$.}

\medskip\noindent {\bf Proof.} Let us consider the complex
extension of $\mathfrak{g}$. We can construct a Chevalley basis
following the same procedure as in the case $\R^{8}$ (see
\ref{app1}). We first define \be H=\ri \what{Y}_3,\quad
E_{+}=\frac12(\what{Y}_1+\ri\what{Y}_2),\quad
E_{-}=\frac12(-\what{Y}_1+\ri\what{Y}_2) \ee with commutation
relations \be [H,E_{+}]=2E_{+},\quad [H,E_{-}]=-2E_{-},\quad
[E_{+},E_{-}]=H . \ee Using these, we define new matrices, which
are linear combinations of the elements defined above:
\begin{align}
\mathcal{H}_i=&(E_{ii}-E_{i+1,i+1})\otimes H,\quad i=1,\ldots, n-1\nonumber\\
\mathcal{H}_n= &E_{nn}\otimes H \nonumber\\
\mathcal{E}_{\pm,j}^{\ell}=&E_{jj}\otimes E_{\pm},\quad j=1,\ldots ,n\label{che}\\
\mathcal{E}_{\pm,jk}^{s,1}=&\pm\frac12(E_{jk}-E_{kj})\otimes I_4+\frac{1}{2}(E_{jk}+E_{kj})\otimes H,\quad 1\le j<k\le n\nonumber\\
\mathcal{E}_{\pm,jk}^{s,2}=&(E_{jk}+E_{kj})\otimes E_{\pm},\quad 1\le j<k\le n.\nonumber
\end{align}

It turns out that, as can be checked by an explicit computation,
the matrices \be \mathcal{H}_i,\quad
\mathcal{E}_{\pm,j}^{\ell},\quad \mathcal{E}_{\pm,jk}^{s,1},\quad
\mathcal{E}_{\pm,jk}^{s,2} \ee form a basis of the Lie algebra
$C_n$ (in the Cartan notation) in a $4n$-dimensional
representation. The matrices $\mathcal{H}_i$ ($i=1,\ldots,n$) are
a basis of a Cartan subalgebra, which we will denote as
$\mathfrak{h}$; the matrices $\mathcal{E}_{\pm,j}^{\ell}$ are the
root vectors corresponding to the long roots, and the
$\mathcal{E}_{\pm,jk}^{s,r}$ to the short ones. The details of the
computation of the commutation relations, and in particular the
determination of the root system, which show that this is the Lie
algebra $C_n$, are given in Appendix \ref{app2}. This concludes
the proof of Theorem 4. \EOP

\medskip\noindent {\bf Proof of Theorem 3 (conclusion).} We can now
conclude the proof of Theorem 3; for this it is necessary to come
back considering $\mathfrak{L} = \mathfrak{su}(2) \oplus
\mathfrak{g} \subset \mathfrak{so}(4n)$. The complex extension of
the orthogonal algebra $\mathfrak{so}(4n)$, is $D_{2n}$ in the
Cartan notation. The maximal subgroups of the classical groups
were classified by Dynkin in \cite{Dy52}, and the result we need
is  that $A_1 \oplus C_{n}$ is a maximal subalgebra of the Lie
algebra $D_{2n}$, which is in agreement with our results in
Theorems 4 and 5.

In fact, the fundamental representation $(10\cdots0)$, in the
highest weight notation, of $D_{2n}$ is irreducible when
restricted to the subalgebra $A_1\oplus C_{n}$, as \be D_{2n}\to
A_1\oplus C_n,\quad (1\overbrace{0\cdots0}^{2n-1})\to
(1)\oplus(1\overbrace{0\cdots0}^{n-1}), \ee and decomposes, when
restricted to $A_1$, into $2n$ copies of the spin 1/2
representation (i.e. $(1)$ in the highest weight notation), \be
D_{2n}\to A_1,\quad (1\overbrace{0\cdots0}^{2n-1})\to 2n(1); \ee
and, when restricted to $C_n$, into the sum of two copies of the
fundamental representation of $C_n$ (we showed this fact by an
explicit computation in Appendix \ref{app1} for the case
$\R^{8}$): \be D_{2n}\to C_{n},\quad
(1\overbrace{0\cdots0}^{2n-1})\to 2(1\overbrace{0\cdots0}^{n-1}).
\ee

Since $\mathfrak{g}$ is a semisimple compact Lie algebra, we
finally have the complete structure of the algebra $\mathfrak{L}$,
i.e. $\mathfrak{L}=\mathfrak{su}(2)\oplus \mathfrak{sp}(n)$, as
claimed. \EOP

\medskip\noindent {\bf Remark 9.} Our discussion shows, as
mentioned in passing, that the invariance algebra
$\mathfrak{L}_n=\mathfrak{su}(2)\oplus \mathfrak{g}\subset
\mathfrak{so}(4n)$ is actually a maximal subalgebra of
$\mathfrak{so}(4n)$. The elements in $\mathfrak{su}(2)$ correspond
to regular invariance transformations, and those in $\mathfrak{g}=
\mathfrak{sp}(n)$ to the strong invariance ones. \EOR

\medskip\noindent {\bf Remark 10.} We should note that although the
invariance algebras are isomorphic for different quaternionic
structures on $\R^{4n}$, their realizations are not the same and
depend on the different quaternionic structures (in particular, on
the different orientations they may have). We have seen in the
case $\R^{8}$ how they can be constructed and the differences
among them. The construction follows essentially the same lines in
the general case. \EOR

\medskip\noindent {\bf Remark 11.} The representation of
$\mathfrak{g}=\mathfrak{sp}(n)$ is complex reducible, and then,
there exists a matrix $P$ which transforms the $4n\times 4n$
matrices into a $2n\times 2n$ block diagonal form. Each block
corresponds to the fundamental representation (of dimension $2n$)
of $C_{n}$. The explicit computation is made for $\R^8$ in
Appendix \ref{app1}, but it cannot be easily generalized. \EOR

\medskip\noindent {\bf Remark 12.} The appearance of the (compact)
symplectic algebra $\mathfrak{sp}(n)$ is not a surprising result
in this context. In fact, it can be identified with the Lie
algebra $\mathfrak{sl}(n,\mathbf{H})$ of quaternionic $n\times n$
matrices with purely imaginary trace \cite{Ba02}. The symplectic
group $\mathrm{Sp}(2n)$ can be realized in terms of quaternions as
a subgroup of the general linear group
$\mathrm{GL}(n,\mathbf{H})$. \EOR

\section{Conclusions}
\label{sec:conclusions}

There is a natural notion of {\it equivalent \HK structures} on a
\HK manifold $(\V,g;J_1,J_2,J_3)$; the \quaternionic (or
hypersymplectic) transformations $\Phi : V \to V$ are those which
preserve the Riemannian metric $g$ and which map the hypercomplex
structure $\{ J_\alpha \}$, and hence the associated
hypersymplectic structure $\{ \omega_\alpha \}$, into an
equivalent one. In the case where the complex structures
$J_\alpha$, and hence the associated symplectic structures
$\omega_\alpha$, are individually invariant under the
transformation $\Phi$, one says that $\Phi$ is strongly
hypersymplectic. It is clear that \HS (and strongly
hypersymplectic) maps form a Lie group.

In this paper we have investigated  adopting a fully explicit approach
the continuous group of \quaternionic (hypersymplectic) transformations for the real
spaces $\R^{4n}$ equipped with the Euclidean metric $g = I_{4n}$,
obtaining a complete classification for their Lie algebra and
hence the connected component of the identity in the group (the
full group is then recovered by taking into account
transformations which permute the different four dimensional
blocks, and possibly other discrete maps).

In particular, we have shown (Theorem 5) that the algebra of
strongly hypersymplectic maps, also called the strong invariance
algebra, is  $\mathfrak{g}_n = \mathfrak{sp}(n)$. As for the full
invariance algebra $\mathfrak{L}_n$, this is always the direct sum
of the strong invariance one and of the regular invariance algebra
(the algebra of transformation mapping the hyperk\"ahler structure
into an equivalent one, different from the original one); we have
shown that the regular invariance algebra is given by
$\mathfrak{su}(2)$, hence (Theorem 3) $\mathfrak{L}_n =
\mathfrak{su} (2) \oplus \mathfrak{sp} (n)$

 As already mentioned in the Introduction, these results are not
new, being known in the differential geometric literature devoted to
\hK and quaternionic manifolds \cite{Jo00}. In this context they were
obtained by rather abstract methods, so that the contribution of this
paper lies in that the proofs are fully explicit and make use of
elementary linear algebra plus Cartan's classification of simple
Lie algebras and Dynkin's classification of maximal subgroups of simple Lie groups.

Our proofs used the standard real quaternionic representation for
SU(2), see Lemma 1, as well as some other Lemmas (see Lemmas 2, 3
and 4 here). In the Euclidean case it turns out we have to deal
with the possibility of different orientations. The Lemmas proved
in this paper show that \HK structures in $(\R^{4n},I_{4n} )$
corresponding to different orientations are conjugated in
$\mathrm{O}(4n)$ and hence lead to isomorphic groups and algebras;
thus one has to deal with a single case (say with fully positive
orientation) for each dimension. As for the study of this single
case, the identification of the regular invariance algebra has
been rather straightforward, while for identifying the strong
invariance algebra we resorted to some general results from the
theory of Lie algebras.

The result discussed here  should be seen as an equivalent in \HK geometry of the
familiar identification of the symplectic group in standard
symplectic geometry. These results also have a relevance in
connection with hyperhamiltonian dynamics \cite{GM02,MT03} and
hence of its physical applications \cite{GR08,GR12} (as well as
its applications in the theory of integrable systems
\cite{GM03a,Ga11}).

The present results only apply to Euclidean spaces; it should be stressed
again that this setting suffices to describe physically relevant cases and equations,
such as the Pauli and the Dirac ones \cite{GR08};  also, most of the physically
relevant \hK manifolds are obtained as quotients (via a momentum map-like construction due
to Hitchin {\it et al.} \cite{HK87}) of standard $\R^{4n}$ \hK manifolds.
Moreover, the fact we were able to fully classify the invariance algebra in this case is
encouraging in view of the treatment of more general cases. In
particular, we have recently been able to fully describe the \HK
structure in Taub-NUT spaces \cite{GR12}; it would be quite
natural to attempt a classification of \quaternionic maps for
these, and hence a classification of Taub-NUT manifolds up to
equivalence of \HK structures.

\section*{Acknowledgements}  MAR was supported by the Spanish Ministry of Science and
Innovation under project FIS2011-22566. GG is supported by the
Italian MIUR-PRIN program under project 2010-JJ4KPA. This article
was started in the course of visits of GG at Universidad
Complutense of Madrid and of MAR at Universit\`a  degli Studi  di Milano.

\appendix

\section{Identification of the strong invariance algebra in the 8-dimensional Euclidean space.}
\label{app1}

We describe the complex Lie algebra $C_2$ using the fundamental
4-dimensional representation. After an appropriated choice of a
basis  of this algebra, we will check that the commutation table
is the same as the one we can compute for the 10-dimensional
subalgebra of the algebra \eqref{basis}. We finally identify the
corresponding real form.

\subsection*{The Lie algebra $C_2$.}

The 4-dimensional fundamental representation of $C_2$ can be defined by the matrices:

\begin{align*}
&X_1=\begin{pmatrix}
 1 & 0 & 0 & 0 \\
 0 & -1 & 0 & 0 \\
 0 & 0 & -1 & 0 \\
 0 & 0 & 0 & 1
\end{pmatrix}
,\quad
X_2=\begin{pmatrix}
0 & 0 & 0 & 0 \\
0 & 1 & 0 & 0 \\
0 & 0 & 0 & 0 \\
0 & 0 & 0 & -1
\end{pmatrix},
\\
&X_3=\begin{pmatrix}
 0 & -1 & 0 & 0 \\
 0 & 0 & 0 & 0 \\
 0 & 0 & 0 & 0 \\
 0 & 0 & 1 & 0
\end{pmatrix}
,\quad
X_4=\begin{pmatrix}
 0 & 0 & 0 & 0 \\
 -1 & 0 & 0 & 0 \\
 0 & 0 & 0 & 1 \\
 0 & 0 & 0 & 0
\end{pmatrix}
,\quad
X_5=\begin{pmatrix}
 0 & 0 & 0 & 0 \\
 0 & 0 & 0 & 1 \\
 0 & 0 & 0 & 0 \\
 0 & 0 & 0 & 0
\end{pmatrix},
\quad
 X_6=\begin{pmatrix}
 0 & 0 & 0 & 0 \\
 0 & 0 & 0 & 0 \\
 0 & 0 & 0 & 0 \\
 0 & 1 & 0 & 0
\end{pmatrix}
,\\ &
X_7=\begin{pmatrix}
 0 & 0 & 0 & 1 \\
 0 & 0 & 1 & 0 \\
 0 & 0 & 0 & 0 \\
 0 & 0 & 0 & 0
\end{pmatrix}
,\quad
X_8=\begin{pmatrix}
 0 & 0 & 0 & 0 \\
 0 & 0 & 0 & 0 \\
 0 & 1 & 0 & 0 \\
 1 & 0 & 0 & 0
\end{pmatrix},
\quad
X_9=\begin{pmatrix}
 0 & 0 & 1 & 0 \\
 0 & 0 & 0 & 0 \\
 0 & 0 & 0 & 0 \\
 0 & 0 & 0 & 0
\end{pmatrix}
,\quad
X_{10}=\begin{pmatrix}
 0 & 0 & 0 & 0 \\
 0 & 0 & 0 & 0 \\
 1 & 0 & 0 & 0 \\
 0 & 0 & 0 & 0
\end{pmatrix},
\end{align*}
satisfying
\be X_i^TJ_4+J_4X_i=0,\quad i=1,\ldots,10,\quad J_4= \begin{pmatrix}
 0 & I_2 \\
 -I_2 & 0
\end{pmatrix}  .
\ee The two first matrices $X_1, X_2$, are a basis of a Cartan
subalgebra related to the two simple roots $\alpha_1$ and
$\alpha_2$. The rest of matrices corresponds to the root spaces:
$\alpha_1$, $-\alpha_1$, $\alpha_2$, $-\alpha_2$,
$\alpha_1+\alpha_2$, $-\alpha_1-\alpha_2$, $2\alpha_1+\alpha_2$
and $-2\alpha_1-\alpha_2$. For our purposes it is more convenient
to use the basis
\begin{align*}
&\wt{X}_1=\frac{1}{2}X_1,\quad \wt{X}_2=\frac{1}{2}X_1+X_2,\quad
\wt{X}_3=-X_9,\quad \wt{X}_4=X_{10},\quad \wt{X}_5=X_5\quad
\wt{X}_6=X_6,\nonumber \\  &\wt{X}_7=\frac{1}{\sqrt{2}}X_8,\quad
\wt{X}_8=\frac{1}{\sqrt{2}}X_4,\quad
\wt{X}_9=-\frac{1}{\sqrt{2}}X_7,\quad
\wt{X}_{10}=-\frac{1}{\sqrt{2}}X_3.
\end{align*}

The commutation table is written in Table \ref{algC2}.
\begin{table}[ht]
\begin{tabular}{c|cccccccccc}
& $\wt{X}_1$ &$\wt{X}_2$ &$\wt{X}_3$ &$\wt{X}_4$ &$\wt{X}_5$ &$\wt{X}_6$ &$\wt{X}_7$ &$\wt{X}_8$ &$\wt{X}_9$ &$\wt{X}_{10}$ \\
\hline
 $\wt{X}_1$ &0 & 0 & $\wt{X}_3$ & $-\wt{X}_4$ & $-\wt{X}_5$ & $\wt{X}_6$ & 0 & $-\wt{X}_8$ & 0 & $\wt{X}_{10}$ \\
 $\wt{X}_2$ &0 & 0 & $\wt{X}_3$ & $-\wt{X}_4$ & $\wt{X}_5$ & $-\wt{X}_6$ & $-\wt{X}_7$ & 0 & $\wt{X}_9$ & 0 \\
 $\wt{X}_3$ &$-\wt{X}_3$ & $-\wt{X}_3$ & 0 & $-\wt{X}_1-\wt{X}_2$ & 0 & 0 & $-\wt{X}_10$ & $\wt{X}_9$ & 0 & 0 \\
 $\wt{X}_4$ &$\wt{X}_4$ & $\wt{X}_4$ & $\wt{X}_1+\wt{X}_2$ & 0 & 0 & 0 & 0 & 0 & $-\wt{X}_8$ & $\wt{X}_7$ \\
 $\wt{X}_5$ &$\wt{X}_5$ & $-\wt{X}_5$ & 0 & 0 & 0 & $-\wt{X}_1+\wt{X}_2$ & $-\wt{X}_8$ & 0 & 0 & $\wt{X}_9$\\
$\wt{X}_6$  &$-\wt{X}_6$ & $\wt{X}_6$ & 0 & 0 & $\wt{X}_1-\wt{X}_2$ & 0 & 0 & $-\wt{X}_7$ & $\wt{X}_{10}$ & 0 \\
$\wt{X}_7$  &0 & $\wt{X}_7$ & $\wt{X}_10$ & 0 & $\wt{X}_8$ & 0 & 0 & $-\wt{X}_4$ & $\wt{X}_2$ & $\wt{X}_6$ \\
 $\wt{X}_8$ &$\wt{X}_8$ & 0 & $-\wt{X}_9$ & 0 & 0 & $\wt{X}_7$ & $\wt{X}_4$ & 0 & $\wt{X}_5$ & $\wt{X}_1$ \\
$\wt{X}_9$ &0 & $-\wt{X}_9$ & 0 & $\wt{X}_8$ & 0 & $-\wt{X}_{10}$ & $-\wt{X}_2$ & $-\wt{X}_5$ & 0 & $-\wt{X}_3$ \\
 $\wt{X}_{10}$ &$-\wt{X}_{10}$ & 0 & 0 & $-\wt{X}_7$ & $-\wt{X}_9$ & 0 & $-\wt{X}_6$ & $-\wt{X}_1$ & $\wt{X}_3$ & 0
\end{tabular}
\caption{Commutation table of $C_2$ in the   basis
$\wt{X}_i$.\label{algC2}}
\end{table}

\subsection*{The strong invariance algebra of the positively oriented
standard quaternionic structure.}

If we consider the 10-dimensional subalgebra in \eqref{basis} with basis ($W_i=Q_{i+3}$)
\begin{align} \label{algquat}
Q_{1,2,3}= \begin{pmatrix}\what{Y}_{\alpha}&0 \\ 0& 0\end{pmatrix}  ,\quad
Q_{4,5,6}= \begin{pmatrix}  0&0 \\ 0& \what{Y}_{\alpha}\end{pmatrix}  ,\quad
Q_{7,8,9}= \begin{pmatrix} 0 & \what{Y}_{\alpha} \\ \what{Y}_{\alpha}& 0\end{pmatrix}  ,\quad
Q_{10}= \begin{pmatrix}0 & I_4 \\ -I_4 & 0\end{pmatrix}  ,
\end{align}
we can easily compute the adjoint representation  and its Killing form:
\be
B=
\begin{pmatrix}
 -12 I_6   \\
   & -24 I_4
\end{pmatrix}  .
\ee Since this quadratic form is non degenerated, it corresponds
to a semisimple subalgebra and, since it is negative definite, it
is compact. We construct its complex extension and change the
basis to:
\begin{align*}
& \wt{Q}_1=\frac{\ri}{2}(Q_3-Q_6),\quad
\wt{Q}_2=\frac{\ri}{2}(Q_3+Q_6),\quad \wt{Q}_3=\frac{1}{2}(Q_1+\ri
Q_2),\quad
\wt{Q}_4=\frac{1}{2}(Q_1-\ri  Q_2),\\
& \wt{Q}_5=-\frac{1}{2}(Q_4+\ri Q_5),\quad
\wt{Q}_6=\frac{1}{2}(Q_4-\ri Q_5),\quad
\wt{Q}_7=-\frac{1}{2 \sqrt{2}}(Q_7-\ri Q_8),\\
& \wt{Q}_8=\frac{1}{2 \sqrt{2}}(\ri Q_9-Q_{10}),\quad
\wt{Q}_9=-\frac{1}{2 \sqrt{2}}(Q_7+\ri Q_8),\quad
\wt{Q}_{10}=-\frac{1}{2 \sqrt{2}}(\ri Q_9+Q_{10}).
\end{align*}

It is straightforward to check that the commutation table in this
basis is exactly the same as in Table \ref{algC2}, and then both
algebras are isomorphic, that is the complex extension of the
algebra constructed with the matrices \eqref{algquat} is $C_2$.

We can easily construct a Chevalley basis in terms of the
quaternionic expressions:
\begin{align*}
&H_1=\ri(Q_3-Q_6),\quad
H_2=\ri Q_6,\\
&E_{\alpha_1}=\frac12(Q_{10}+\ri Q_9),\quad
E_{-\alpha_1}=\frac12(-Q_{10}+\ri Q_9),\\
&E_{\alpha_2}=\frac12(Q_4+\ri Q_5),\quad
E_{-\alpha_2}=\frac12(-Q_4+\ri Q_5),\\
&E_{\alpha_1+\alpha_2}=\frac12(Q_7+\ri Q_8),\quad
E_{-\alpha_1-\alpha_2}=\frac12 (-Q_7+\ri Q_8),\\
&E_{2\alpha_1+\alpha_2}= \frac12(Q_1+\ri Q_2),\quad
E_{-2\alpha_2-\alpha_2}= \frac12(-Q_1+\ri Q_2),
\end{align*}
that is
\begin{align*}\label{mat}
 &H_1=\ri \begin{pmatrix}\what{Y}_3 &0\\ 0 &-\what{Y}_3\end{pmatrix}  ,\quad
H_2=\ri  \begin{pmatrix} 0 & 0 \\ 0 & \what{Y}_3\end{pmatrix}  ,\\
& E_{\alpha_1}=\frac12 \begin{pmatrix} 0 & I_4+\ri \what{Y}_3\\ -I_4+\ri\what{Y}_3 & 0\end{pmatrix}  ,\quad
E_{-\alpha_1}=\frac12 \begin{pmatrix} 0 & -I_4+\ri \what{Y}_3\\ I_4+\ri\what{Y}_3 & 0\end{pmatrix}  ,\\
& E_{\alpha_2}=\frac12 \begin{pmatrix} 0 & 0\\ 0 & \what{Y}_1+\ri\what{Y}_2\end{pmatrix}  ,\quad
E_{-\alpha_2}=\frac12 \begin{pmatrix} 0 & 0\\ 0 & -\what{Y}_1+\ri\what{Y}_2\end{pmatrix}  ,\\
& E_{\alpha_1+\alpha_2}=\frac12 \begin{pmatrix} 0 & \what{Y}_1+\ri \what{Y}_2\\ \what{Y}_1+\ri \what{Y}_2 & 0\end{pmatrix}  ,\quad
E_{-\alpha_1-\alpha_2}=\frac12  \begin{pmatrix}0 & -\what{Y}_1+\ri \what{Y}_2\\ -\what{Y}_1+\ri \what{Y}_2 & 0\end{pmatrix}  ,\\
& E_{2\alpha_1+\alpha_2}= \frac12 \begin{pmatrix}\what{Y}_1+\ri \what{Y}_2 & 0\\ 0 & 0\end{pmatrix}  ,\quad
E_{-2\alpha_1-\alpha_2}= \frac12 \begin{pmatrix}-\what{Y}_1+\ri\what{Y}_2 & 0 \\ 0 & 0\end{pmatrix}
\end{align*}
which will be generalized to the general case $\R^{4n}$
\eqref{che}. Finally, the Lie algebra $C_2$ has no irreducible
8-dimensional representation. However, the matrix \be P=
\frac{1}{\sqrt{2}} \begin{pmatrix}
 1 & 0 & 0 & 0 & 0 & 0 & 1 & 0 \\
 \ri & 0 & 0 & 0 & 0 & 0 & -\ri & 0 \\
 0 & 0 & -1 & 0 & 1 & 0 & 0 & 0 \\
 0 & 0 & -\ri & 0 & -\ri & 0 & 0 & 0 \\
 0 & -1 & 0 & 0 & 0 & 0 & 0 & -1 \\
 0 & -\ri & 0 & 0 & 0 & 0 & 0 & \ri \\
 0 & 0 & 0 & 1 & 0 & -1 & 0 & 0 \\
 0 & 0 & 0 & \ri & 0 & \ri & 0 & 0
\end{pmatrix}
\ee converts the matrices $\Phi_i$ of \eqref{algquat} into a
block-diagonal form: \be P^{-1}\Phi_iP= \begin{pmatrix}X_i & 0\\ 0
& X_i\end{pmatrix}  , \ee showing in an explicit way that they are
a reducible representation of $C_2$, which can be decompose into
the sum of two copies of the fundamental representation of $C_2$
with dimension 4.

Among the real forms of $C_2$ there is only one which is compact,
$\mathfrak{sp}(2)$, which is the algebra we were looking for.

\section{Commutation relations of the Lie algebra $C_n$}\label{app2}

It is straightforward to check the commutation relations proving
that the basis \be \mathcal{H}_i,\quad
\mathcal{E}_{\pm,j}^{\ell},\quad \mathcal{E}_{\pm,jk}^{s,1},\quad
\mathcal{E}_{\pm,jk}^{s,2} \ee is a Chevalley basis corresponding
to $C_n$. First, the elements in
$\mathfrak{h}=\{\mathcal{H}_1,\ldots,\mathcal{H}_n\}$ clearly
commute (they are block diagonal matrices and the blocks are equal
to 0 or to $H$): \be [\mathcal{H}_i,\mathcal{H}_j]= 0,\quad
i,j=1,\ldots ,n. \ee

Second, the matrices $\mathcal{E}_{\pm,j}^{\ell}$,
$\mathcal{E}_{\pm,jk}^{s,1}$, $\mathcal{E}_{\pm,jk}^{s,2}$ have
the correct commutation relations, that is, they are eigenvectors
of the elements in the Cartan subalgebra:
\begin{align}
& [\mathcal{H}_i,\mathcal{E}_{\pm,j}^{\ell}]=\pm 2
(\delta_{ij}-\delta_{i+1,j})\mathcal{E}_{\pm,j}^{\ell},\quad
i=1,\ldots,n-1,\quad j=1,\ldots ,n
\nonumber\\
& [\mathcal{H}_n,\mathcal{E}_{\pm,j}^{\ell}]=\pm 2
\delta_{nj}\mathcal{E}_{\pm,j}^{\ell}, \quad j=1,\ldots ,n
\nonumber\\
&
[\mathcal{H}_i,\mathcal{E}_{\pm,jk}^{s,1}]=\pm(\delta_{ij}-
\delta_{ik}-\delta_{i+1,j}+\delta_{i+1,k})\mathcal{E}_{\pm,jk}^{s,1},
\quad i=1,\ldots,n-1,\quad 1\le j<k\le n
\nonumber\\
&
[\mathcal{H}_n,\mathcal{E}_{\pm,jk}^{s,1}]=\pm(\delta_{nj}-
\delta_{nk})\mathcal{E}_{\pm,jk}^{s,1},
\quad  1\le j<k\le n
\nonumber\\
& [\mathcal{H}_i,\mathcal{E}_{\pm,jk}^{s,2}]= \pm(\delta_{ij}+
\delta_{ik}-\delta_{i+1,j}-\delta_{i+1,k})\mathcal{E}_{\pm,jk}^{s,2},
\quad i=1,\ldots,n-1,\quad 1\le j<k\le n
\nonumber\\
& [\mathcal{H}_n,\mathcal{E}_{\pm,jk}^{s,2}]=\pm(\delta_{nj}+
\delta_{nk})\mathcal{E}_{\pm,jk}^{s,2}, \quad 1\le j<k\le n.\label{comm3}
\end{align}

Let us consider the vector space of the Cartan subalgebra
$\mathfrak{h}$ and the forms defined by: \be
e_j:\mathfrak{h}\to\mathfrak{h},\quad e_j(E_{ii}\otimes
H)=\delta_{ij} \ee and then \be
e_j(\mathcal{H}_i)=e_j(E_{ii}\otimes H-E_{i+1,i+1}\otimes
H)=\delta_{ij}-\delta_{i+1,j},\quad
e_j(\mathcal{H}_n)=e_j(E_{nn}\otimes H)=\delta_{nj}. \ee

Using these roots, the commutation relations \eqref{comm3} can be written as:
\begin{align*}
& [\mathcal{H}_i,\mathcal{E}_{\pm,j}^{\ell}]=\pm 2 e_j(\mathcal{H}_i)\mathcal{E}_{\pm,j}^{\ell},\quad i=1,\ldots,n-1,\quad j=1,\ldots ,n
\\
& [\mathcal{H}_n,\mathcal{E}_{\pm,j}^{\ell}]=\pm 2 e_j(\mathcal{H}_n)\mathcal{E}_{\pm,j}^{\ell}, \quad j=1,\ldots ,n
\\
& [\mathcal{H}_i,\mathcal{E}_{\pm,jk}^{s,1}]=\pm(e_j-e_k)(\mathcal{H}_i)\mathcal{E}_{\pm,jk}^{s,1}, \quad i=1,\ldots,n-1,\quad 1\le j<k\le n
\\
& [\mathcal{H}_n,\mathcal{E}_{\pm,jk}^{s,1}]=\pm(e_j-e_k)(\mathcal{H}_n)\mathcal{E}_{\pm,jk}^{s,1}, \quad  1\le j<k\le n
\\
& [\mathcal{H}_i,\mathcal{E}_{\pm,jk}^{s,2}]= \pm(e_j+e_k)(\mathcal{H}_i)\mathcal{E}_{\pm,jk}^{s,2}, \quad i=1,\ldots,n-1,\quad 1\le j<k\le n
\\
& [\mathcal{H}_n,\mathcal{E}_{\pm,jk}^{s,2}]=\pm (e_j+e_k)(\mathcal{H}_n)\mathcal{E}_{\pm,jk}^{s,2}, \quad 1\le j<k\le n.
\end{align*}
That is, the root system is \be 2e_i,\quad \pm e_j\pm e_k,\quad
i,j,k=1,\ldots,n \ee which is the root system of the Lie algebra
$C_n$. This ends the proof of Theorem 5.2.

The commutation relations between a root vector and that
associated to the opposite root are:
\begin{align*}\label{roots}
 [\mathcal{E}_{+,i}^{\ell},\mathcal{E}_{-,i}^{\ell}]=&E_{ii}\otimes H=\mathcal{H}_i+\mathcal{H}_{i+1}+\cdots+\mathcal{H}_{n},\quad  n=1,\ldots ,n
\\
 [\mathcal{E}_{+,jk}^{s,1},\mathcal{E}_{-,jk}^{s,1}]=&(E_{jj}-E_{kk})\otimes H=\mathcal{H}_j+\mathcal{H}_{j+1}+\cdots+\mathcal{H}_{k-1}, \quad 1\le j<k\le n
\\
 [\mathcal{E}_{+,jk}^{s,2},\mathcal{E}_{-,jk}^{s,2}]=& (E_{jj}+E_{kk})\otimes H=\mathcal{H}_j+\mathcal{H}_{j+1}+\cdots+\mathcal{H}_{k-1}+2\mathcal{H}_k+\cdots+2\mathcal{H}_n, \;  \\ & 1\le j<k\le n.
\end{align*}
We can easily identified a set of simple roots. The root vectors
associated to the simple short roots are: \be
\mathcal{E}_{+,12}^{s,1},\ldots,\mathcal{E}_{+,n-1,n}^{s,1},\quad
[\mathcal{E}_{+,i,i+1}^{s,1},\mathcal{E}_{-,i,i+1}^{s,1}]=\mathcal{H}_i,\quad
i=1,\ldots,n-1 \ee and the root vector associated to the simple
long root is \be \mathcal{E}_{+,n}^{\ell},\quad
[\mathcal{E}_{+,n}^{\ell},\mathcal{E}_{-,n}^{\ell}]=\mathcal{H}_n.
\ee

\end{document}